\documentclass[manuscript]{acmart}
\usepackage{courier}
\usepackage[ruled,vlined]{algorithm2e}
\usepackage{paralist}
\usepackage[shortlabels]{enumitem}
\usepackage{tabularx}
\usepackage{balance}
\usepackage{multirow}
\usepackage{adjustbox}
\usepackage{multicol}
\usepackage{pbox}
\usepackage{mathtools}
\usepackage{subcaption}
\usepackage{algorithmic}
\usepackage{url}
\usepackage{setspace}	
\usepackage{verbatim}
\usepackage{float}
\usepackage[normalem]{ulem}
\usepackage{xspace}
\usepackage{ifthen}
\usepackage{bold-extra}
\usepackage{hyperref}
\usepackage{pifont}
\usepackage{tikz}
\usepackage{xcolor}
\usepackage{listings}

\AtBeginDocument{%
  }

\setcopyright{none}
\copyrightyear{2025}
\acmYear{2025}
\acmDOI{XXXXXXX.XXXXXXX}

\acmJournal{TOSEM}
\acmVolume{0}
\acmNumber{0}
\acmArticle{0}
\acmMonth{0}

\begin{document}
\input{macro.tex}

\title{Reassessing Code Authorship Attribution in the Era of Language Models}

\author{Atish Kumar Dipongkor}
\email{atish.kumardipongkor@ucf.edu}
\affiliation{%
  \institution{University of Central Florida}
  \city{Orlando}
  \state{Florida}
  \country{USA}
}

\author{Ziyu Yao}
\email{ziyuyao@gmu.edu}
\affiliation{%
 \institution{George Mason University}
 \city{Fairfax}
 \state{Virginia}
 \country{USA}}

\author{Kevin Moran}
\email{kpmoran@ucf.edu}
\affiliation{%
  \institution{University of Central Florida}
  \city{Orlando}
  \state{Florida}
  \country{USA}
}

\renewcommand{\shortauthors}{Dipongkor, et al.}

\begin{abstract}
The study of Code Stylometry, and in particular Code Authorship Attribution (CAA), aims to analyze coding styles to identify the authors of code samples. CAA has been illustrated to be an important component of automating software engineering (SE) tasks such as bug triaging, fault localization, and test prioritization. In addition, CAA is also important in cybersecurity and software forensics for addressing copyright disputes and detecting plagiarism. Past techniques for CAA tend to leverage hand-crafted code-related features that typically carry limitations that prevent proper authorship characterization and lead to sensitivities to adversarial attacks.
 
Recently, transformer-based Language Models (LMs) have shown remarkable efficacy across a range of SE tasks, and in authorship attribution for natural language in the NLP domain. However, their effectiveness in CAA is not well understood. As such, we conduct the first extensive empirical study applying two larger state-of-the-art code LMs and five smaller code LMs to the task of CAA on six diverse datasets that encompass 12K code snippets written by 463 developers. Furthermore, we perform an in-depth quantitative and qualitative analysis of our studied models' performance on CAA using established interpretability techniques. Our results illustrate important aspects of the behavior of LMs in understanding stylometric code patterns.
\end{abstract}



\maketitle

\section{Introduction}
\label{sec:Intro}

The study of Code Authorship Attribution (CAA) centers around the analysis of coding styles, including coding structure, naming conventions, and formatting styles, to identify the authors of code samples~\cite{abuhamad2018large,alsulami2017source}. CAA plays a vital role in assisting with the automation of key software engineering (SE) tasks related to maintenance and software quality. For example, in automated bug triaging past research has aimed to use code authorship as means to build profiles of expertise for developers to more effectively triage bugs~\cite{dipongkor2023comparative}. However, authorship information in software repositories may often be absent or inaccurate due to processes such as code review or pair programming~\cite{bogomolov2021authorship,anvik2006should}. As such, CAA can assist in these tasks by disambiguating and accurately providing more reliable authorship information. Similar benefits of CAA hold true across other SE tasks such as fault localization, developer profiling and workload balancing~\cite{seeberger2006developers}, Security and Compliance Audits~\cite{bird2011don,rahman2011ownership}, and test prioritization~\cite{yin2011fixes}. \revision{More specifically CAA can aid with \emph{Bug Triaging and Fault Localization} by identifying the most likely author of a buggy code segment, which accelerates the debugging process and reduces maintenance overhead~\cite{anvik2006should,fritz2010degree}; \emph{Developer Profiling and Workload Balancing} by analyzing developer coding styles, which can help assign tasks based on expertise levels, optimizing team productivity and reducing risk~\cite{seeberger2006developers}; \emph{Security and Compliance Audits} by attributing code authorship can detect unauthorized contributions or potential insider threats, supporting secure software practices~\cite{bird2011don,rahman2011ownership}; and finally \emph{Testing Efficiency}: wherein stylometric patterns can inform targeted testing strategies, such as prioritizing tests for modules written by less experienced developers. This can reduce test suite size and accelerate validation cycles~\cite{thongtanunam2016revisiting}. Given the role that CAA, and the stylometric identification of coding patterns more broadly, have across various stages of the software development lifecycle, its importance is clear.}

Beyond development, CAA is also critical in application domains such as cybersecurity and software forensics for addressing copyright disputes, detecting plagiarism, and supporting criminal prosecutions~\cite{Ropgen,caliskan2015anonymizing}. Consequently, this area of research is becoming increasingly popular~\cite{caliskan2015anonymizing}. However, manually verifying or identifying authors is tedious and time-consuming; these tasks are susceptible to human errors and inconsistencies, especially in large software systems~\cite{frantzeskou2006effective}. These difficulties tend to arise because such systems often involve numerous authors, who often exhibit nuanced overlaps in coding styles (e.g., due to the presence of style guidelines or linters) that can make differentiation among contributors difficult~\cite{liu2021practical}.

Considering the above challenges, researchers have proposed approaches using Machine Learning (ML) and Deep Learning (DL) techniques to automate tasks related to CAA~\cite{abuhamad2018large,abuhamad2021large,abuhamad2019code}. However, these techniques suffer from two key limitations. Namely, they typically rely on \textit{hand-crafted} feature engineering and they are susceptible to \textit{adversarial attacks}, wherein authors attempt to avoid identification through purposeful code perturbations. Hand-crafted features are problematic because there are many different aspects that can define coding style (e.g., length, whitespace, declaration usage, identifier style) and manually defining features to properly capture a range of different coding styles is challenging.
Prior models have also been shown to be susceptible to adversarial attacks
~\cite{quiring2019misleading}. Given these limitations, further research is needed to understand how to best design practical, robust techniques for carrying out CAA.

Recently, transformer-based \LLMsL{} (\LLMsS{}) for code have shown remarkable effectiveness across several SE tasks including but not limited to Code Clone Detection, Vulnerability Detection, and Software Documentation~\cite{raychev2014code,du2023extensive}. Much of the work that adapts \LLMsS{} to SE tasks originated in the field of Natural Language Processing (NLP). In fact, stylometry for \textit{natural language} is well studied within the field of NLP, where \LLMsS{} have been illustrated to excel at the task~\cite{fourkioti2019language,fabien2020bertaa}. However, \LLMsS{} have not been well studied for tasks related to \textit{code} stylometry. We argue that it is worthwhile to understand the application of \LLMsS{} to code stylometry for two main reasons. First, \LLMsS{} are typically trained on massive datasets, spanning both code and natural language, meaning that they have likely observed a wide array of coding styles, and hence may perform well at picking up on the nuanced patterns that define authorship. Second, models fine-tuned for \CAAS{}  can take advantage of the automated feature engineering capabilities of \LLMsS{}, making use of rich code embeddings for CAA. 

As such, in this paper, we carry out the first comprehensive, large-scale empirical study applying LMs to the task of CAA, with the intent of examining in what settings LMs excel at the task, their limitations across datasets of varying code composition, and their robustness to adversarial code transformations. Additionally, we aim to gain a better understanding of \textit{why} we observe the model performance reported in our study through an in-depth analysis of LM behavior using machine learning interpretability techniques.

Our study context encompasses 12K code snippets gathered from six different sources, one of which is a novel dataset not explored in prior work. These datasets are highly diverse in terms of coding styles, with contributions from 463 developers. We fine-tune two larger code \LLMsS{}, \CodeLlama{}~\cite{codellama} and \DeepSeek{}~\cite{deepseekcoder}, and five smaller code \LLMsS{} including \CodeBERT{}~\cite{codebert}, \ContraC{}~\cite{contrabert}, \ContraG{}~\cite{contrabert}, \GraphCodeBERT{}~\cite{graphcodebert} and \UnixCoder{}~\cite{unixcoder}, and evaluate their effectiveness in CAA across our studied datasets. More importantly, we also study the robustness of these models to adversarial attacks, by applying carefully crafted code perturbations to our dataset samples that have been shown effective at deceiving prior work. Finally, we conduct both qualitative and quantitative analysis by adapting well-known interpretability techniques to our model and problem context in order to better understand \textit{why} we observe the model performance captured in our study. Our analysis captures several significant findings that further our understanding of how \LLMsS{} for code perform on tasks related to code stylometry, and point toward important directions for future investigation. 
The key findings from our study are as follows:

\begin{enumerate}[label=\textbf{(\arabic*)}]

\item \textbf{Larger and Smaller \LLMsS{} are Effective in Different Settings}:
Larger \LLMsS{} like \CodeLlama{} have significant capacity, and when properly fine-tuned for the \CAAS{} task, tend to perform well on \textit{multilingual, imbalanced datasets with shorter code snippets}. Smaller fine-tuned models tend to perform well on \textit{balanced, monolingual datasets}.

\item \textbf{\LLMsS{} Outperform Prior Techniques by Learning More Separable Author Embeddings}:
The advantage of \LLMsS{} over \PbNN{}~\cite{bogomolov2021authorship} is explained by embedding-space geometry rather than dataset size. Fine-tuned \LLMsS{} separate intra- and inter-author cosine-similarity distributions with substantially higher Jensen-Shannon Divergence (\JSD{})~\cite{bevendorff2020divergence} than \PbNN{} (e.g., \textit{0.992 vs. 0.924} on GitHub-Java; \textit{0.666 vs. 0.181} on \LeetCode{}), and this separability predicts attribution accuracy across datasets. Pre-trained code embeddings encode authorship signal that path-based embeddings learned from scratch fail to capture.

\item \textbf{Stylometric Variation Influences Attribution Performance}:
A larger number of training samples does not always translate to better performance. For instance, in a dataset we derive from \LeetCode{}~\cite{leetcode}, \textit{C++ (62.19\%) had the most samples but achieved only 64\% accuracy}, whereas \textit{C\# (2.02\%) had fewer samples but attained 94\% accuracy}. This discrepancy arises due to the reduced stylometric variation in C++, where many developers solve the same problem in a similar style, making authorship attribution harder.

\item \textbf{Misattribution is Concentrated on Style-Overlap and Short Snippets}:
Across all models, misclassified samples have intra- and inter-author similarity distributions that overlap almost completely (\JSD{} = 0.016)---different authors converge on identical idioms in algorithmically constrained problems. Independently, snippets below 200 tokens carry insufficient stylometric signal.

\item \textbf{\LLMsS{} Exhibit Orthogonal Stylometric Understanding}:
Different \LLMsS{} are able to learn \textit{distinct, non-overlapping stylistic features} to attribute authorship. The \textit{top 30\% most important features} (identified using \xaiMethod{}~\cite{sundararajan2017axiomatic}, validated via necessity and sufficiency) play a crucial role in this process, and different models tend to learn different ``important'' features, motivating future work on combining different models.

\item \textbf{\LLMsS{} Are Less Vulnerable to Adversarial Attacks Than Prior ML/DL Techniques}:
Although adversarial success rates vary across models, all \LLMsS{} except \CodeLlama{}---which we were only able to partially fine-tune via LoRA---are more robust against adversarial attacks than \PbNN{} (\textit{\PbNNAdversarialSuccessRate{}}), which also exhibits the highest category-wise success rate across every transformation family.

\item \textbf{Adversarial Robustness Depends on Feature-Utilization Strategy, Not Just Model Size}:
Using \xaiMethod{} over 14 AST token categories, we find smaller \LLMsS{} (\CodeBERT{}, \GraphCodeBERT{}, \ContraC{}, \ContraG{}---each \smallLLMParameters{} parameters) distribute attribution roughly uniformly across many categories, while larger \LLMsS{} (\UnixCoder{} at \UnixCoderParameters{}, \DeepSeek{} at \DeepSeekParameters{}, \CodeLlama{} at \CodeLlamaParameters{}) and \PbNN{} concentrate attribution on a few categories (e.g., \PbNN{} on \textit{preprocessor} and \textit{comments}; \UnixCoder{} on \textit{identifiers}; \CodeLlama{} on \textit{keywords}). Since each adversarial transformation perturbs only a subset of categories, distributed-attention models retain discriminative signal in untouched categories and degrade less. Robustness is therefore governed by \textit{what features the model relies on}, not parameter count alone.

\item \textbf{Contrastive Pre-Training Forces Structural Feature Learning}:
\ContraC{} and \ContraG{}---pre-trained with an InfoNCE objective over semantic-preserving augmentations (variable/function rename, statement reorder, dead-code insertion)---show substantially lower adversarial success rates than their non-contrastive base models (\CodeBERT{}, \GraphCodeBERT{}) on exactly the attacks that mirror those augmentations. The mechanism: contrastive training pulls semantically equivalent variants together in embedding space, pushing the model to rely on \textit{structural and syntactic} features rather than surface identifiers.

\item \textbf{Adversarial Changes that Modify Higher Number of Tokens are More Successful}:
The most effective adversarial transformation rules involve \textit{removing comments, eliminating unused code, adding print/log statements, modifying functions, and altering control flow}. These transformations change a larger portion of the code, making them more effective in deceiving \LLMsS{}.

\end{enumerate}

\section{Problem Statement and Research Questions}

\CAAS{} involves analyzing a piece of code to determine who wrote it when the author is unknown.  
In this paper, as in prior work~\cite{bogomolov2021authorship}, we formulate this problem as a multi-class classification problem, wherein a model takes a snippet of code as input and predicts one of a known set of authors. In this paper, we evaluate the effectiveness of \LLMsS{} on CAA using the following research questions (RQs):

\RQ{1}: \textbf{How effective are \LLMsS{} for \CAAS{}?} This RQ aims to evaluate the effectiveness of \LLMsS{} in identifying the author of a given code snippet across diverse scenarios. Specifically, we assess their performance across different programming languages, code snippet sizes, and number of potential unique authors. To provide a comprehensive analysis, we compare the performance of \LLMsS{} with an existing DL-based technique, examining their relative strengths and limitations in CAA.

\RQ{2}: \textbf{What stylistic features of code do \LLMsS{} learn when applied to \CAAS{}?} \CAAS{} relies on the assumption that each developer has a unique coding style that models can learn to distinguish. While \LLMsS{} may achieve high performance in CAA, it remains unclear whether they genuinely capture distinct stylistic features for each author or rely on spurious correlations. If \LLMsS{} identify orthogonal stylistic features for each author, it would indicate that they are learning author-specific patterns. However, if the important features overlap significantly across authors, it would suggest that the models are making attributions based on shared structures rather than individual stylistic traits. Understanding this distinction is essential to ensure that \LLMsS{} provide reliable and interpretable authorship predictions rather than merely optimizing for high accuracy through non-meaningful signals. This RQ aims to investigate whether \LLMsS{} truly capture orthogonal stylistic features in \CAAS{}, by assessing their interpretability and robustness.

\RQ{3}: \textbf{How robust are \LLMsS{} against adversarial attacks for the \CAAS{} task?} This RQ investigates the vulnerability of \LLMsS{} in \CAAS{} when exposed to adversarial modifications. Since \LLMsS{} may rely on superficial stylistic patterns, they can be susceptible to attacks such as variable renaming, code refactoring, and statement reordering, which preserve original program logic while altering surface-level features. We aim to assess whether \LLMsS{} capture deep, meaningful stylistic patterns or merely rely on brittle heuristics. Understanding their robustness will provide insights into potential weaknesses and inform the development of more resilient attribution models.

\section{Studied Datasets}

\begin{table}[tb]
    \caption{Summary of the Datasets used for \CAAS{}}
    \label{tbl:datasets}
    \setlength{\tabcolsep}{2pt}
    \centering
    \scalebox{0.90}{
    \begin{tabular}{l|c|c|c|c}
    \hline
    \textbf{Dataset} & \textbf{Authors (A)} & \textbf{Samples (S)} & \textbf{Min S per A} & \textbf{Max S per A} \\ \hline
    \GCJCPP{}          & 20                   & 160                  & 8                    & 8                    \\ \hline
    \GCJJava{}         & 74                   & 2202                 & 24                   & 48                   \\ \hline
    \GCJPython{}         & 66                   & 660                  & 10                   & 10                   \\ \hline
    \GithubC{}         & 66                   & 1916                 & 10                   & 86                   \\ \hline
    \GithubJava{}      & 39                   & 2667                 & 11                   & 583                  \\ \hline
    \LeetCode{}         & 198                  & 4753                 & 10                   & 197                  \\ \hline
    \end{tabular}}
    \end{table}

To answer the above RQs, we utilized six datasets collected from GitHub, Google Code Jam (GCJ), and LeetCode, whose details are provided in Table~\ref{tbl:datasets}. The first five datasets are employed in existing studies~\cite{Ropgen,yang2022natural}, while we collected the last one ourselves. The datasets \GCJJava{}~\cite{Ropgen}, \GCJCPP{}~\cite{Ropgen}, and \GCJPython{}~\cite{yang2022natural} originate from the GCJ challenge. The other two datasets, \GithubC{}~\cite{Ropgen} and \GithubJava{}~\cite{Ropgen}, are sourced from GitHub. We compiled the last dataset from LeetCode. For this dataset, we scraped public solutions to various LeetCode problems. For instance, there are public solutions~\cite{two-sum-solution} to the \textit{Two Sum}~\cite{two-sum-description} problem. For each LeetCode problem, we collected all public solutions and submitted them to the platform to check for correctness. If the solution was accepted, it was included in our dataset. This method ensured the validity of the solutions, as not all public solutions were correct. \revision{
Note that we have included Table~\ref{tbl:leetcode-details-stat} in our paper appendix that presents a comprehensive breakdown of the LeetCode dataset across six programming languages. The dataset contains 4,753 total samples from 198 authors. Note that the author counts and samples-per-author statistics in Table~\ref{tbl:leetcode-details-stat} are computed \textit{within} each language: because many authors contribute in more than one language, these per-language counts sum to more than 198. Correspondingly, the Min and Max samples-per-author reported for \LeetCode{} in Table~\ref{tbl:datasets} are per-author totals across \textit{all} languages, which is why they differ from the within-language values in Table~\ref{tbl:leetcode-details-stat}. C++ dominates with 2,956 samples (62.19\%) from 142 authors, followed by Java with 1,180 samples (24.83\%) from 87 authors, and Python with 322 samples (6.77\%) from 29 authors. The remaining languages---JavaScript, C\#, and Ruby---constitute 6.21\% of the dataset combined. The samples-per-author (S/A) statistics reveal significant imbalance within each language: the maximum S/A ranges from 13 (Ruby) to 150 (C++), while standard deviations of S/A span from 4.99 (Ruby) to 23.86 (C\#), indicating highly skewed author contributions. Notably, C\# shows the highest relative variability with only 4 authors contributing between 1 and 56 samples each, while C++ exhibits the largest absolute spread with authors contributing between 1 and 150 samples.
} Finally, we contributed this dataset to evaluate \LLMsS{} in the following scenarios:

\textbf{(1) Multilingual Scenario}: As displayed in Table~\ref{tbl:datasets}, existing datasets are monolingual, containing samples from a single programming language, which often does not replicate the real world. In industry and open-source projects, developers often use multiple programming languages. For instance, a project may be written in Java but utilize Python for testing or JavaScript for web interfaces. This multilingual aspect is crucial for real-world \CAAS{} applications.

\begin{figure}
    \centering
    \includegraphics[width=\textwidth]{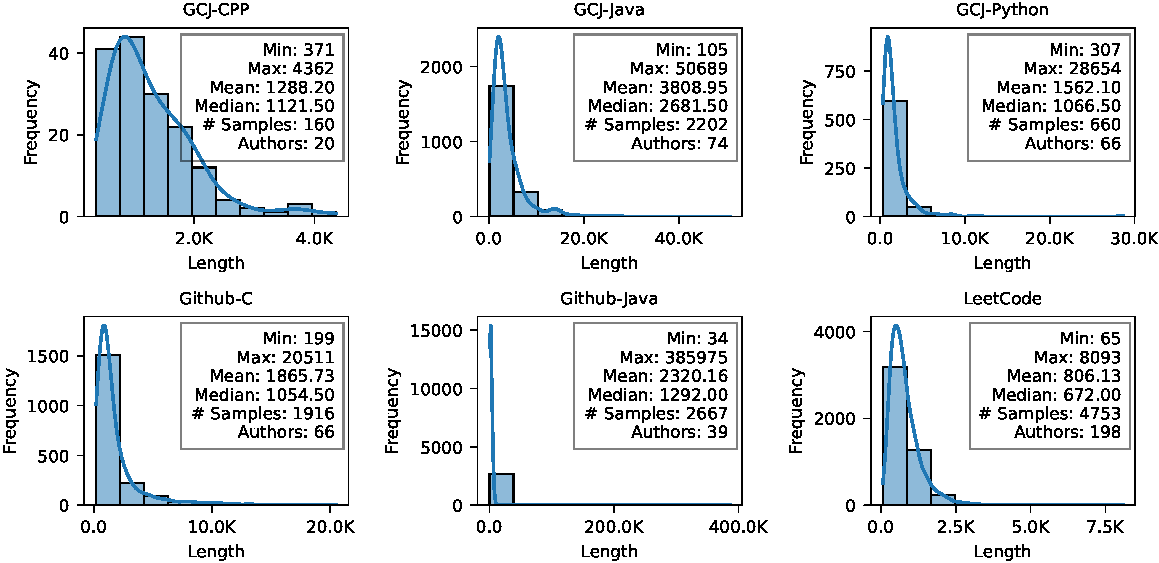}
    \caption{Example Distributions of Code Sample Size in two Studied Datasets.}
    \Description{Box plots comparing the distribution of code sample sizes across two studied datasets, showing that LeetCode samples are shorter than existing datasets.}
    \label{fig:dataset-distribution}
\end{figure}

\textbf{(2) Smaller Code Size and Imbalanced Data}: As shown in Figure~\ref{fig:dataset-distribution}, the code samples in the existing datasets are relatively longer than those in the \LeetCode{} dataset. As such, we cannot evaluate the performance of \LLMsS{} on small code samples using the existing datasets. However, in real-world scenarios, it is common to encounter small code snippets, such as those found in bug reports or code reviews. Therefore, we included the \LeetCode{} dataset to evaluate the performance of \LLMsS{} on small code samples. Additionally, existing datasets are mostly balanced, with each author contributing a similar number of samples. However, in real-world scenarios, the number of samples per author can vary significantly. For instance, some authors may contribute only a few lines of code, while others may contribute entire modules or libraries. Therefore, we included the \LeetCode{} dataset to evaluate the performance of \LLMsS{} on datasets with a large number of authors and varying numbers of samples per author.

\textbf{(3) Less Stylometric Variation among Authors}: In open-source or industry projects, developers often follow similar coding styles and conventions, making it challenging to distinguish between authors based on stylistic features. This is particularly true for large projects with many contributors, where the codebase is often subject to strict style guidelines and code reviews. While solving \LeetCode{} problems, users often need to use the best data structure and algorithm to avoid time and space complexity issues. Therefore, there is less stylometric variation among authors.

\textbf{(4) Adversarial Attack Scenario}: While attacking \LLMsS{} or any model, there are two main constraints that need to be considered: (1) the attack should not change the original logic of the code, and (2) adversarial samples should be plausible~\cite{quiring2019misleading}. Prior CAA datasets do not include test cases, which makes it challenging to fulfill the first constraint. In our \LeetCode{} dataset, we can test the first constraint by submitting the adversarial samples to \LeetCode{} and checking if they are accepted. We discuss the second constraint for the \LeetCode{} dataset in the threats to validity section.

\section{Empirical Study Methodology}

\subsection{\RQ{1}: \LLMsS{} Effectiveness in CAA}
To answer \RQ{1} and \RQ{2}, we fine-tune seven pre-trained \LLMsS{} for code including \CodeBERT{}~\cite{codebert}, \ContraC{}~\cite{contrabert}, \ContraG{}~\cite{contrabert}, \GraphCodeBERT{}~\cite{graphcodebert}, \UnixCoder{}~\cite{unixcoder}, \DeepSeek{}~\cite{deepseekcoder}, and \CodeLlama{}~\cite{codellama} on the \CAAS{} task. \LLMsS{} for code fall into several categories based on their architecture, pretraining objectives and number of parameters. This study focuses on evaluating \textit{open-source} LMs for CAA tasks. GPT-4/4o-like models are closed models, and their weights are not publicly accessible, making them difficult/impossible to fine-tune for CAA and unsuitable for interpretability analyses. Our interpretability methods require access to internal model representations, which is not possible with GPT-4-based APIs or other commercial models.

Due to computational resource constraints, it was not reasonably practical to include \textit{all} open-source models in our experiments. Instead, we selected representative models from all major transformer architectures—including encoder-only, decoder-only, and encoder-decoder—to ensure broad coverage. Specifically, we included \DeepSeek{} (1.3B) and \CodeLlama{} (7B) as large-scale models to assess how model capacity influences performance. Finally, our selection considered varied pre-training objectives (e.g., masked language modeling, causal language modeling, contrastive learning), as these strategies are known to impact downstream performance. Our experimental design therefore balances architectural diversity, scale, and pre-training strategies, providing insights into how these factors influence CAA performance. The properties and attributes of these \LLMsS{} are summarized in Table~\ref{tab:code_models}. In the evaluation section, we try to understand the effectiveness and generalization of different types of \LLMsS{} for \CAAS{} by comparing their performance on different datasets.

\begin{table*}[tb]
    \centering
    \caption{Comparison of Studied Code Language Models}
    \scalebox{0.80}{
    \begin{tabular}{p{8cm} | p{8cm}}
        \toprule
        \textbf{Model (Parameter Size, Short Description, Pretraining Data)} & \textbf{Details (Type, Layers, Input Limit, Pretraining Objective)} \\ \hline
        \textbf{\CodeLlama{} (7B)} - LLaMA-based model for code generation \& completion, trained on 500B tokens of code data from GitHub.
        & \textbf{Type:} Decoder-Only, \textbf{Layers:} 32, \textbf{Max Input:} 16,384 tokens, \textbf{Pretraining:} Causal LM \\ \hline

        \textbf{\DeepSeek{} (1.3B)} - Large instruct-tuned model for code generation, trained on 2T tokens from multilingual data from GitHub. 
        & \textbf{Type:} Decoder-Only, \textbf{Layers:} 24, \textbf{Max Input:} 4,096 tokens, \textbf{Pretraining:} Causal LM + Instruction Tuning \\ \hline

        \textbf{\UnixCoder{} (223M)} - Unified pretraining framework for code understanding \& generation, trained on CodeSearchNet, GitHub and StackOverflow.
        & \textbf{Type:} Encoder-Decoder, \textbf{Layers:} 24, \textbf{Max Input:} 1,024 tokens, \textbf{Pretraining:} MLM (Masked Language Modeling) + PLM (Permuted Language Modeling) + RTD (Replaced Token Detection) \\ \hline

        \textbf{\CodeBERT{} (125M)} - Bimodal model trained on source code and NL, using CodeSearchNet (Python, Java, JavaScript, PHP, Ruby, Go) + NL.
        & \textbf{Type:} Encoder-Only, \textbf{Layers:} 12, \textbf{Max Input:} 512 tokens, \textbf{Pretraining:} MLM + RTD \\ \hline

        \textbf{\GraphCodeBERT{} (125M)} - Extends \CodeBERT{} with structure-aware learning, trained on CodeSearchNet + Data Flow Graphs.
        & \textbf{Type:} Encoder-Only, \textbf{Layers:} 12, \textbf{Max Input:} 512 tokens, \textbf{Pretraining:} MLM + Edge Prediction \\ \hline

        \textbf{\ContraC{} (125M)} - Robust version of \CodeBERT{} using contrastive learning, trained on CodeSearchNet + Augmented Data.
        & \textbf{Type:} Encoder-Only, \textbf{Layers:} 12, \textbf{Max Input:} 512 tokens, \textbf{Pretraining:} MLM + Contrastive Learning \\ \hline

        \textbf{\ContraG{} (125M)} - Robust version of \GraphCodeBERT{} with contrastive pretraining, trained on CodeSearchNet + Data Flow + Augmented Data.
        & \textbf{Type:} Encoder-Only, \textbf{Layers:} 12, \textbf{Max Input:} 512 tokens, \textbf{Pretraining:} MLM + Contrastive Learning \\
        \hline
    \end{tabular}}
    \label{tab:code_models}
\end{table*}

Figure~\ref{fig:dataset-distribution} and Table~\ref{tab:code_models} show that most samples exceed the input size of our smaller studied \LLMsS{}. As such, we split larger code samples into smaller chunks, ensuring that each chunk fits within the model's input limit while retaining author information. This approach applies to all \LLMsS{} except \CodeLlama{}. \revision{For \DeepSeek{}, GPU memory constraints led us to limit chunks to 2500 tokens rather than its maximum input size of 4,096 tokens.}

To fine-tune all encoder-only \LLMsS{}, we adopt the standard sequence classification approach~\cite{devlin2018bert}, which involves adding a classification layer to the embedding of the [CLS] token. For \CodeLlama{} and \DeepSeek{}, we use the embedding of the last token as input to the classification layer. Fine-tuning is performed using the cross-entropy loss function, with the Adam optimizer~\cite{kingma2014adam} for weight updates. For all models, we set the maximum number of epochs to 20 and apply early stopping with a patience of 3. All experiments are conducted using the Hugging Face library~\cite{huggingface}.

In addition to this setup, we conduct hyperparameter tuning for the learning rate and batch size using a grid search approach, following an existing CAA study~\cite{bogomolov2021authorship}. We evaluate three learning rates (5e-05, 3e-05, and 2e-05) and two batch sizes (16 and 32), as these values have been shown to be effective for fine-tuning \LLMsS{} on downstream tasks~\cite{devlin2018bert}. This results in six configurations, each evaluated using \textit{10-fold cross-validation} for all datasets except \GCJCPP{}, where we use \textit{8-fold cross-validation} due to there being only 8 samples per author. \revision{We apply this grid to every model listed in Table~\ref{table:hyperparams}, including \DeepSeek{}. Because \DeepSeek{} cannot accommodate a micro-batch of 16 or 32 on a single GPU, we realize each grid batch size through gradient accumulation, using a per-device micro-batch of 4 with 4 or 8 accumulation steps so that the effective batch size matches the grid value.} We select the best hyperparameter set based on the average F1 score across all folds. \revision{In order to bolster the reproducibility of our work, we have included Table~\ref{table:hyperparams} in our appendix that provides a list of the best hyperparameters for each model.}

Our experiments reveal that the optimal hyperparameters vary across models and datasets. However, a learning rate of 3e-05 with a batch size of 16 is the most frequently selected configuration, accounting for 9 of the 36 model--dataset pairs in Table~\ref{table:hyperparams}; a batch size of 16 is selected in 25 of the 36 cases. Fine-tuning is conducted on a single NVIDIA H100 GPU with 80GB of memory. \revision{For \DeepSeek{}, the grid search selects a batch size of 16 on every dataset, with a learning rate of 2e-05 on three datasets, 3e-05 on two, and 5e-05 on one.} Due to memory constraints, we do not apply grid search to \CodeLlama{}. Instead, we tune only its learning rate while fixing the batch size at 2, and a learning rate of 8e-5 performs best across all datasets. \revision{
For \CodeLlama{}, we could not fine-tune all 7B parameters due to GPU memory constraints. Therefore, we employed LoRA~\cite{hu2022lora} to fine-tune the projection layers in the self-attention mechanism. Following the empirical recommendations from Hu~\etal{}~\cite{hu2022lora}, we applied LoRA specifically to the query (Wq) and value (Wv) projection layers, as their analysis demonstrated that this combination achieves optimal performance compared to other configurations (e.g., adapting individual matrices or all four attention matrices). We set rank $r=32$ and $\alpha=64$, resulting in approximately 16.78M trainable parameters (0.24\% of the model). While LoRA has been shown to match or exceed full fine-tuning performance across various tasks~\cite{hu2022lora}, we note that our approach is particularly effective on balanced datasets with high stylometric variation, as evidenced by the results in Table~\ref{tab:caa_results}. Additionally,  we conducted experiments with two other configurations: (rank=$\left\{8,16\right\}$ and alpha=$\left\{16,32\right\}$). As shown in Table~\ref{tbl-lora-config}, all three configurations yield nearly identical performance, with accuracy differences within $\pm$1 percentage point.  
}

For evaluating the effectiveness of \LLMsS{} on the \CAAS{} task, we use \Accuracy{}, \Precision{}, \Recall{}, and \FScore{} as metrics. \Accuracy{} is widely used in the literature~\cite{Ropgen,yang2022natural} to evaluate the effectiveness of baselines on the \CAAS{} task. However, \Accuracy{} is biased towards majority classes, meaning if a model predicts the majority author correctly while failing on minority authors, it may still report a high accuracy. Therefore, we consider the other metrics as well to provide a more balanced evaluation. To ensure a robust assessment of model performance, we employ \textit{K-fold cross-validation}, which mitigates biases from a single train-test split and provides a more stable performance estimate across different subsets of the dataset. This approach helps assess the generalization of \LLMsS{} by ensuring that all authors contribute to both training and testing phases. As we mentioned above, we split the code samples into smaller chunks to fit the input size of \LLMsS{}. To obtain the final prediction for a given code sample, we aggregate the predictions of its chunks by summing up their scores and selecting the author with the highest total score as the final prediction.

We employ the Mann-Whitney U-test~\cite{mcknight2010mann} to ascertain whether a statistically significant difference in performance exists between \LLMsS{}. When comparing any two \LLMsS{}, the null hypothesis ($H_0$) posits that there is no significant difference in the average performance ($\mu_0 = \mu_1$) of the two. Conversely, the alternative hypothesis suggests a significant difference in the average performance ($\mu_0 \ne \mu_1$) of the two \LLMsS{}. Then we calculate the U-statistic, which is based on the difference in the summed ranks between the two models' accuracy scores across $k$-folds. If the p-value associated with the U-statistic is less than the level of significance ($\alpha$), we reject the null hypothesis and conclude that a significant difference exists between the two models' performance. Otherwise, we fail to reject the null hypothesis and conclude that no significant difference exists.

Finally, we compare the effectiveness of \LLMsS{} with a state-of-the-art (SOTA) DL-based technique \PbNN{}~\cite{bogomolov2021authorship}. \PbNN{} (Path-based Neural Network) is a language-agnostic authorship attribution model designed to analyze source code. It is based on the code2vec~\cite{alon2018general} model, which transforms code into numerical representations by extracting AST (Abstract Syntax Tree) paths. We use the implementation of \PbNN{} provided by the authors~\cite{bogomolov2021authorship} and follow the same hyperparameters as in their paper. We use the same evaluation metrics as for \LLMsS{} to compare the performance of \PbNN{} with \LLMsS{}. Note that while other CAA baselines exist, e.g., RoPGen~\cite{Ropgen}, these are often language specific, and given the multilingual nature of most \LLMsS{}, we opted to compare against a similarly multi-lingual model in \PbNN{}.

\subsection{\RQ{2}: What Patterns Do \LLMsS{} Learn in CAA?}
\label{sec:rq2-methodology}

Some \LLMsS{} may achieve exceptionally high \Accuracy{}, \Precision{}, \Recall{}, and \FScore{} in \CAAS{}. However, high performance alone does not confirm that these models are genuinely learning meaningful patterns. It is possible that they rely on spurious correlations rather than identifying distinct stylistic features for each author. Therefore, we investigate whether \LLMsS{} capture orthogonal stylistic features of code during authorship attribution.

\textbf{Qualitative Analysis}: By orthogonal stylistic features, we mean that the model relies on a distinct and non-overlapping set of stylistic features for each author. If \LLMsS{} truly learn unique stylistic patterns per author, then their decision-making process should be based on features that are specific to individual authors and uncorrelated across different authors. Conversely, if the models achieve high accuracy but rely on overlapping or non-distinctive features, this would suggest that \LLMsS{} may be making predictions by chance, reducing the reliability of their authorship attribution capabilities. Here, we implicitly assume that effective models for CAA will maximize linear separability of the learned feature representations. The rationale is that linearly separable feature spaces allow for clear, interpretable decision boundaries between authors' styles, which is desirable for both performance and interpretability.

To answer this research question, we employ an interpretability method, called \xaiMethod{}~\cite{sundararajan2017axiomatic}. This method assigns an importance score to each feature in the input, indicating its contribution to the model's prediction. By analyzing these feature attributions, we can determine whether the model focuses on distinct stylistic aspects for different authors. While interpretability methods such as \xaiMethod{} provide insights into feature importance, it is critical to assess whether these interpretations are faithful—that is, whether they accurately reflect the model's actual decision-making process. To do this, we evaluate faithfulness using two key metrics: \Necessity{} and \Sufficiency{}. These metrics are well-established in the interpretability literature~\cite{tuan2021local,rai2023explaining,atanasova2024diagnostic,alvarez2018towards} and measure how crucial the identified important features are for the model's predictions.

\Necessity{} quantifies whether removing the most important features (as identified by \xaiMethod{}) significantly increases the model's uncertainty. If \LLMsS{} genuinely rely on these features, removing them should lead to higher perplexity in predictions—i.e., the model should struggle more to make accurate predictions. A higher \Necessity{} value means that removing the identified features significantly increases the model's uncertainty, confirming that these features were crucial for prediction.

\begin{equation}
    \Necessity{} := \exp \left( \frac{1}{m} \sum_{i=1}^{m} \left( - \log P_\theta(y_i | x_{R_i}) + \log P_\theta(y_i | x_i) \right) \right)
\end{equation}

where:
\begin{itemize}
    \item \( m \) is the number of test samples.
    \item \( x_i \) is the original input for the \( i \)-th test sample.
    \item \( x_{R_i} \) is the perturbed input after removing the most important tokens identified by \xaiMethod{}.
    \item \( P_\theta(y_i | x_i) \) is the softmax probability of the predicted class for the full input.
    \item \( P_\theta(y_i | x_{R_i}) \) is the softmax probability of the predicted class after perturbation.
\end{itemize}

\Sufficiency{} measures whether the model can still make accurate predictions using only the important features identified by \xaiMethod{}. If these features truly drive the model's decision-making, then predictions should remain confident, meaning perplexity should stay low.

\begin{equation}
    \Sufficiency{} := \exp \left( \frac{1}{m} \sum_{i=1}^{m} - \log P_\theta(y_i | x_{A_i}) \right)
\end{equation}
    
where:
    \begin{itemize}
        \item \( x_{A_i} \) is the reduced input, containing only the most important tokens identified by an explanation method.
        \item \( P_\theta(y_i | x_{A_i}) \) is the softmax probability of the predicted class using only \( x_{A_i} \).
    \end{itemize}

    To assess whether \LLMsS{} capture orthogonal stylistic features, we analyze \Necessity{} and \Sufficiency{} for the top-k\% important features for each \LLM{}. An increase in \Necessity{} and a decrease in \Sufficiency{} as $k$ increases indicate that the model relies on these features for predictions. Next, we examine whether these features are distinct across authors. If the top-k\% features are largely non-overlapping, it suggests that \LLMsS{} capture orthogonal stylistic patterns unique to each author. Conversely, if significant overlap exists, it implies that the models may be relying on shared structures rather than individual stylistic signals, raising concerns about the reliability of \LLMsS{} for the \CAAS{} task.

\textbf{Quantitative Analysis}: \revision{While the feature overlap analysis using \xaiMethod{} provides insights into token-level importance, it does not fully capture whether different models learn fundamentally different representations of authorship styles. To complement our interpretability analysis and provide stronger quantitative evidence for feature orthogonality, we perform a distributional analysis of code similarities in the learned embedding spaces.
}

\revision{
\emph{Stylometric Variation}---If different language models truly learn orthogonal stylometric features, then the geometric relationships between code samples in their respective embedding spaces should differ substantially. Specifically, we expect that intra-author similarities (code pairs from the same author) should be distinctly separated from inter-author similarities (code pairs from different authors), and this separation pattern should vary across different models, indicating that each model captures unique stylistic dimensions.
}

\revision{\emph{Embedding Generation}---For each fine-tuned model in our 10-fold cross-validation setup, we extract code embeddings from the model's final hidden states. We tokenize each code sample with padding and truncation to the maximum sequence length supported by the corresponding model, then pass it through the model to obtain contextualized representations. To create a single embedding vector for each code sample, we use mean pooling: we average the hidden states across all tokens in the sequence, excluding padding tokens. This is achieved by masking out padding tokens using the attention mask, summing the remaining hidden states, and dividing by the number of actual (non-padding) tokens. This mean-pooling approach provides a holistic representation of the entire code sequence.
}

\revision{\emph{Similarity Computation}---We compute pairwise cosine similarities between all code embeddings within each test fold. For each pair of samples $(i,j)$ where $i < j$, we calculate:}

\begin{equation}
    \textit{similarity}(i,j) = \frac{e_i \cdot e_j}{\|e_i\| \, \|e_j\|}
\end{equation}
\revision{where $e_i$ and $e_j$ are the embedding vectors for samples $i$ and $j$, respectively. We then categorize each similarity score into two groups:}

\begin{itemize}
    \item \revision{Intra-author similarities: Pairs where both samples belong to the same author ($\textit{author}(i) = \textit{author}(j)$).}
    \item \revision{Inter-author similarities: Pairs where samples belong to different authors ($\textit{author}(i) \ne \textit{author}(j)$).}
\end{itemize}

\revision{\emph{Distributional Divergence Analysis} - To quantify the separation between intra-author and inter-author similarity distributions, we calculate the Jensen-Shannon Divergence(\JSD{})~\cite{bevendorff2020divergence}, a symmetric measure of distributional difference. The \JSD{} ranges from 0 (identical distributions) to 1 (completely non-overlapping distributions). A high \JSD{} indicates that the model creates distinctly different similarity patterns for same-author versus different-author code pairs, providing quantitative evidence that the model has learned to separate authors based on stylometric features in its embedding space.}

\subsection{\RQ{3}: Effectiveness of Adversarial Changes}  
Existing literature has explored adversarial attacks on \CAAS{} models in two main categories: targeted and non-targeted attacks~\cite{Ropgen,yang2022natural,quiring2019misleading}. In targeted attacks, the adversary attempts to mislead the model into attributing a code snippet to a \textit{specific incorrect author}. In non-targeted attacks, the goal is to cause the model to make \textit{any incorrect attribution}. Prior studies~\cite{Ropgen,yang2022natural,quiring2019misleading} have shown that non-targeted attacks are more effective for \CAAS{}, achieving higher attack success rates. Therefore, we focus on non-targeted attacks in this study.  

Several methods exist for generating adversarial code samples~\cite{quiring2019misleading,liu2021practical,yang2022natural,tian2023code}. However, these methods have limitations: (1) they are often specific to certain programming languages, or (2) they lack support for a wide range of program transformations. To address these limitations, we leverage a \textit{prompt-based approach} using \GPT{}~\cite{achiam2023gpt} to generate adversarial code samples. We used the \texttt{GPT-4o-2024-08-06} snapshot via OpenAI's API, with experiments conducted in February 2025. While there is the potential for non-determinism with GPT-4o, we did attempt to mitigate this by setting the temperature to 0 and using default deterministic parameters. Our approach consists of the following steps:

\begin{enumerate}
    \item \textbf{Generating adversarial samples:} A code transformation prompt and a code snippet are submitted to \GPT{}.
    \item \textbf{Functionality verification:} The adversarially transformed code is submitted to \LeetCode{} to verify that its functionality remains unchanged.  
    \item \textbf{Transformation verification:} Since \GPT{} is prone to hallucinations, we implemented a parser to verify whether the transformations were performed correctly.  
    \item \textbf{Evaluating \LLMsS{} robustness:} If both verifications are successful, we query the \LLMsS{} with the adversarial samples to assess their resilience to adversarial attacks.  
\end{enumerate}

\begin{figure}
    \centering
    \includegraphics[width=0.9\textwidth]{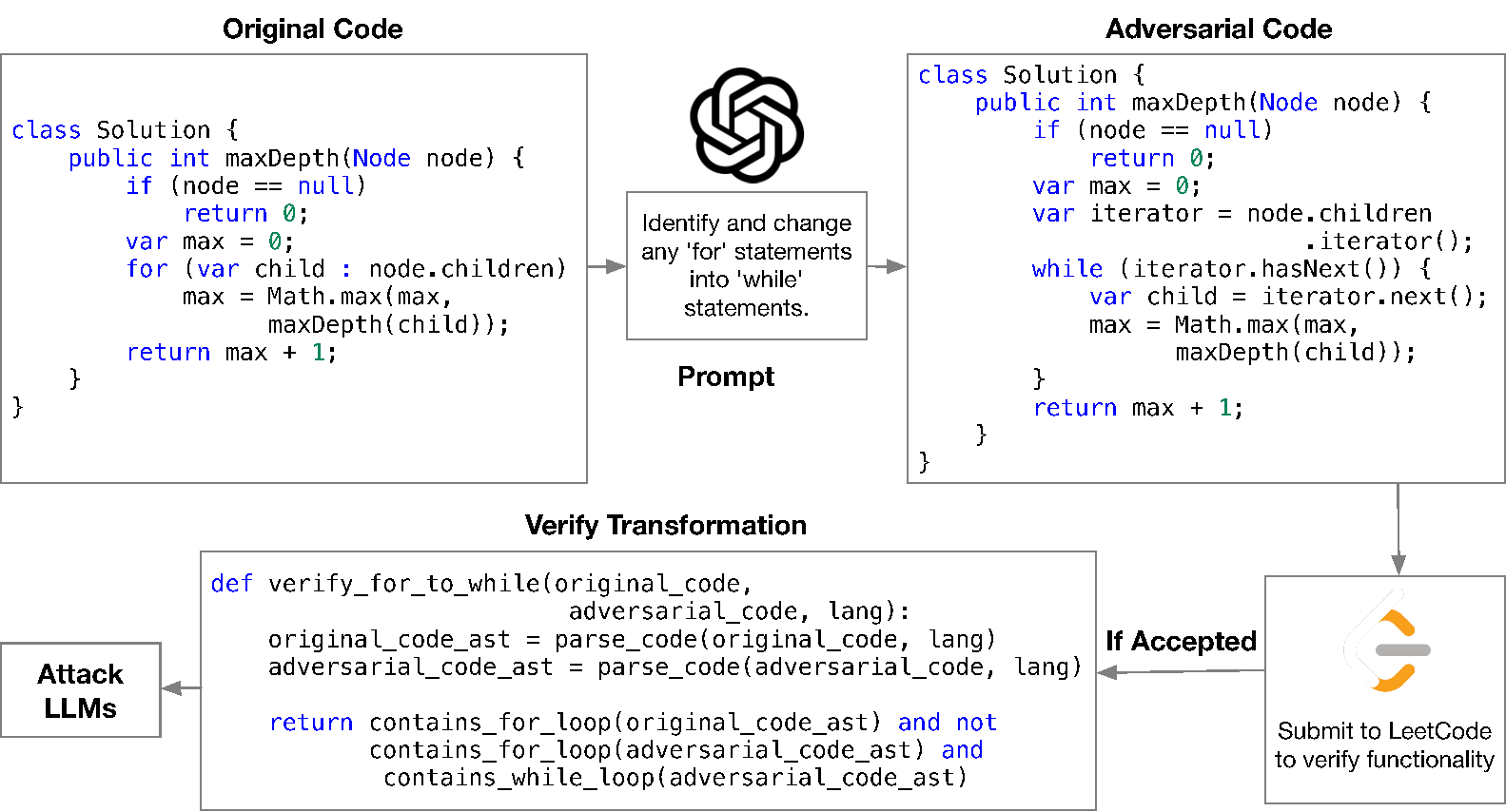}
    \caption{Adversarial Sample Generation and Verification}
    \Description{Flowchart illustrating the adversarial sample generation pipeline: code transformation via GPT-4, functionality verification on LeetCode, transformation verification via parser, and model robustness evaluation.}
    \label{fig:adversarial-attack}
\end{figure}

Figure~\ref{fig:adversarial-attack} illustrates the full adversarial sample generation and verification process. The code transformation rules used in this study are derived from transformation rules established in previous adversarial attacks for \CAAS{}~\cite{liu2021practical,quiring2019misleading,Ropgen}. Below, we present a high-level summary of the code transformation rules \revision{and verification steps} used in this study; full details are given in Appendix~\ref{sec-adv-prompts-rules}.

\begin{enumerate}[label=\textbf{(\Alph*)}]
    \item \textbf{Statement Transformations}: This category includes splitting or combining variable declarations, swapping the order of statements, and modifying the order of independent statements.

    \revision{\noindent \textbf{Verification:} We parse the original and adversarial code and compare their statements. A valid transformation should result in a mismatch between the statement structures of the two programs.}
    
    \item \textbf{Name Transformations}: This category includes rules that involve changing the naming style of variables and functions. For example, converting snake case to camel case or vice versa.

   \revision{\noindent \textbf{Verification:} We extract and compare the identifiers from both the original and adversarial code. A valid transformation should result in a mismatch in the identifier names between the two programs.}

    \item \textbf{Operator Transformations}: This category includes swapping relational operators (`$a > b$' to `$b < a$'), converting conditional statements to inverse operators (`$a < b$' would be `!$(a > b)$'), converting integer literals to expressions (`int b = 2 * 4'), and changing the style of increment or decrement operators (`i++' to `i +=1').

    \revision{\noindent \textbf{Verification:} We extract the AST nodes corresponding to conditional, increment, and decrement operators in both programs. A valid transformation should produce a mismatch in the relevant operators.}

    \item \textbf{Data Transformations}: This category includes rules converting integers to hexadecimal, character literals to ASCII, string variables to character arrays, and boolean literals to integers.

    \revision{\noindent \textbf{Verification:} We analyze the AST nodes related to variable declarations in both programs. A valid transformation should result in a mismatch in the declaration statements.}

    \item \textbf{Loop Transformations}: This category includes rules converting `for' to `while' statements and vice versa.

    \revision{\noindent \textbf{Verification:} We parse loop structures in both programs and compare them. A valid transformation should result in dissimilar loop structures.}

    \item \textbf{Control Flow Transformations}: This category involves converting `if-else' statements to `switch' statements, `switch' statements to `if-else' statements, `if-else' statements to ternary operators, and swapping the statements within `if-else' statements.

    \revision{\noindent \textbf{Verification:} We extract and compare conditional structures from both programs. A valid transformation should result in a mismatch between the control flow structures.}

    \item \textbf{Function Transformations}: This category involves swapping the order of function parameters, adding an extra integer parameter with a default value of zero, creating a new function from a group of statements, and swapping the order of function declarations.
    
    \revision{\noindent \textbf{Verification:} We parse function declarations and invocations from both programs and check for differences in structure. A valid transformation should produce non-identical function signatures or call structures.}

    \item \textbf{Miscellaneous Transformations}: This category includes removing comments, unused code, and adding print/log statements to track variable values.

    \revision{\noindent \textbf{Verification:} We parse both versions of the code and analyze the node types. For comment removal, the original code should contain comment nodes while the adversarial version should not. For unused code elimination, the adversarial version should have fewer AST nodes. Print/log statement insertions are verified using regular expressions.}
\end{enumerate}

Using the above rules, we create 28 code transformation prompts to generate adversarial code samples. Then, each adversarial code sample is verified for functionality and transformation correctness before being used to evaluate the \LLMsS{}' robustness to adversarial attacks. In \revision{Appendix~\ref{sec-adv-prompts-rules}}, we provide a detailed description of each code transformation prompt. We also provide the details of our parser for transformation verification in our \revision{online appendix~\cite{online-appendix}}.

\revision{In addition to our attack based on the semantic-preserving transformations described above, we also attack our baseline models using ALERT~\cite{yang2022natural}, a prominent adversarial attack method for pre-trained models of code. ALERT generates adversarial examples through variable renaming, using the masked language modeling function and contextualized embeddings from the victim model's underlying pre-trained architecture (e.g., \CodeBERT{} or \GraphCodeBERT{}) to produce semantically natural variable substitutes. While ALERT claims to be a black-box attack, it requires substantial knowledge of the victim model's architecture and access to the base pre-trained model's internal functions—making it more accurately characterized as a gray-box approach. In contrast, our attack based on semantic-preserving code transformations is a true black-box attack that requires no knowledge of the victim model's architecture or internals. We leverage GPT-4 with carefully designed code transformation prompts to generate diverse adversarial candidates, followed by semantic equivalence validation to ensure the transformed code preserves the original functionality. Our approach only requires query access to the victim model's predictions, making it more practical and generalizable across model architectures without needing access to their underlying pre-trained components.}

\section{Results}
\subsection{\RQ{1}: Effectiveness of \LLMsS{} for \CAAS{}}
\begin{figure}
    \centering
    \includegraphics[width=\columnwidth]{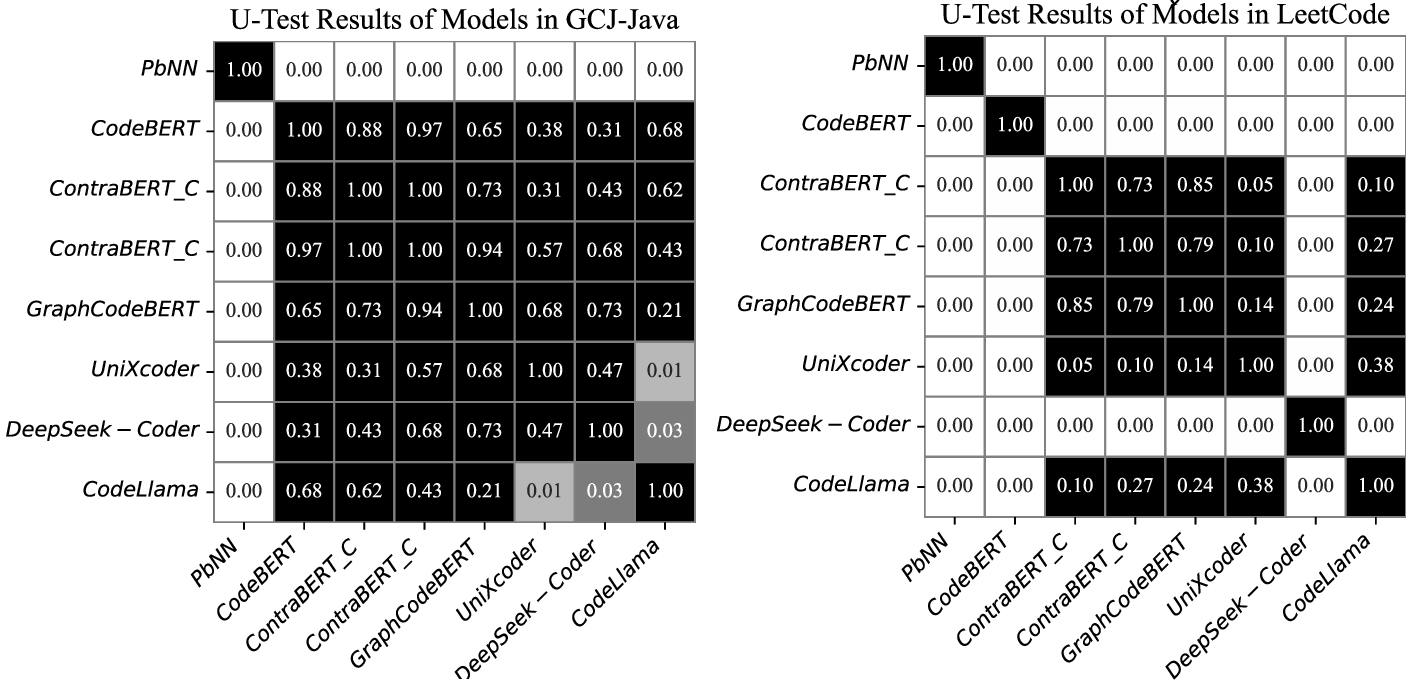}
    \caption{U-test results of Models for GCJ-Java and LeetCode. The black cells indicate that the performance between models is not statistically significant.}
    \Description{Heatmap matrices showing pairwise Mann-Whitney U-test results for GCJ-Java and LeetCode datasets, with black cells indicating non-significant performance differences.}
    \label{fig:model-sig-test}
\end{figure}

Table~\ref{tab:caa_results} presents the results of our \textit{K-fold cross-validation} experiments, including the \Accuracy{}, \Precision{}, \Recall{} and \FScore{} for each model across all datasets. Figure~\ref{fig:model-sig-test} shows the results of the U-test for two datasets. 
The black cells indicate that the performance difference between the two models is not statistically significant since the p-value is greater than the level of significance (0.05). Based on these results, we discuss the effectiveness of \LLMsS{} for \CAAS{}.

 \textbf{SOTA vs \LLMsS{}}: By using the replication package provided by \PbNN{}, we have achieved slightly higher \Accuracy{} than the \Accuracy{} reported for the \GCJCPP{}, \GCJJava{}, \GithubC{} and \GithubJava{} datasets by Li~\etal{}~\cite{Ropgen}. In \PbNN{}'s hyperparameter settings, they ran this model for 10 epochs. From the training dynamics, we observe that its \Accuracy{} increases even after 10 epochs. Therefore, we ran \PbNN{} for 40 epochs with an early stopping patience of 10 during our replication. This is likely the reason for slightly better \Accuracy{}.

\begin{figure}[tb]
    \centering
    \includegraphics[width=0.6\columnwidth]{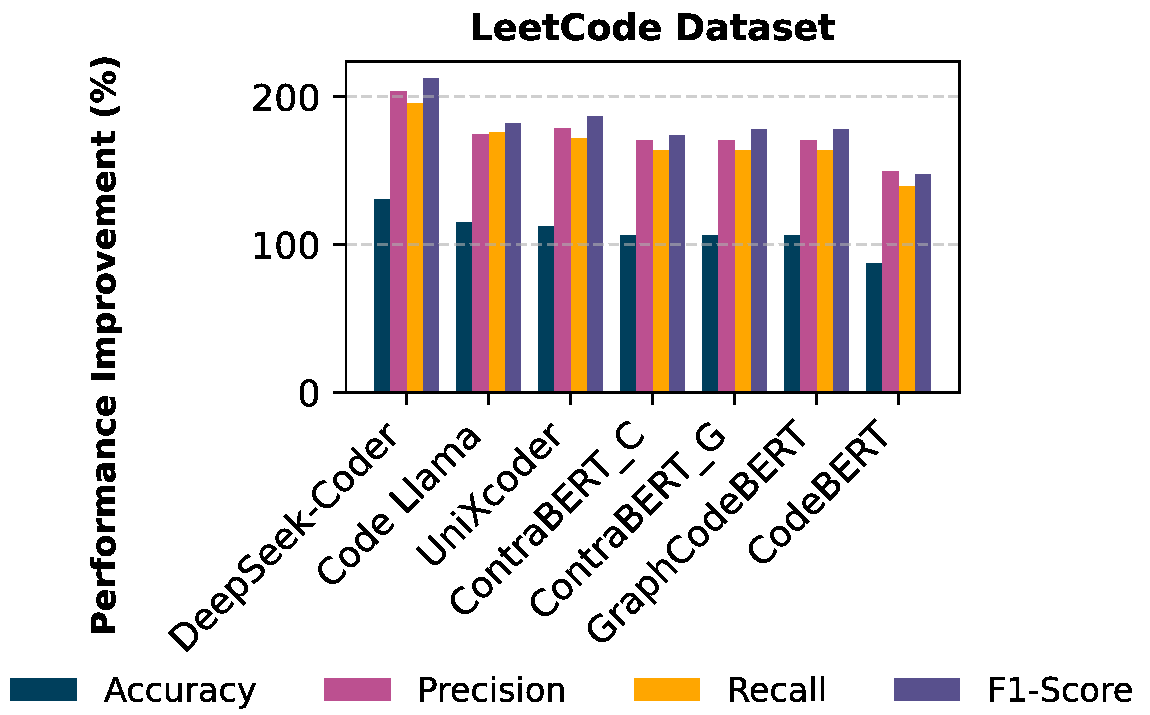}
    \caption{Performance improvement of \LLMsS{} over \PbNN{}}
    \Description{Bar chart showing percentage improvement of language models over \PbNN{} across accuracy, precision, recall, and F1-score metrics for each dataset.}
    \label{fig:performance_improvement_of_llms}
\end{figure}

Figure~\ref{fig:performance_improvement_of_llms} illustrates the percentage improvement of \LLMsS{} over \PbNN{} in terms of \Accuracy{}, \Precision{}, \Recall{}, and \FScore{}. The improvement is computed as:
\begin{equation}
    \frac{(\textit{Metric}_{\textit{LMs}} - \textit{Metric}_{\textit{PbNN}})}{\textit{Metric}_{\textit{PbNN}}} \times 100\revision{\%}
\end{equation}

The results demonstrate a substantial advantage of \LLMsS{} over \PbNN{} across all datasets in \CAAS{}. In particular, the \LeetCode{} dataset exhibits the largest performance gap, where \LLMsS{} achieve over 130\% improvement in \Accuracy{}, up to 204.17\% in \Precision{}, 196\% in \Recall{}, and 213.04\% in \FScore{}. These improvements highlight the limitations of \PbNN{} in handling \textit{multilingual, imbalanced, short code snippets with minimal stylometric variation}, all of which are key characteristics of \LeetCode{}. Across other datasets, \LLMsS{} outperform \PbNN{} by up to 38.03\% in \Accuracy{}, 53.97\% in \Precision{}, 38.03\% in \Recall{}, and 46.97\% in \FScore{}.

\begin{table*}[tb]
    \centering
    \caption{\textit{K-fold cross-validation} performance of \LLMsS{} and baselines on the CAA task. The best results are in \textbf{bold}.}
    \label{tab:caa_results}
    \setlength{\tabcolsep}{0.2pt}
    \scalebox{0.77}{
    \begin{tabular}{l|c|c|c|c|c|c|c|c|c}
    \hline
    Dataset                      & Metric    & \PbNN{} & \CodeBERT{}      & \CodeLlama{}     & \ContraC{} & \ContraG{} & \DeepSeek{}      & \GraphCodeBERT{} & \UnixCoder{}     \\ \hline
    \multirow{4}{*}{\LeetCode{}}    & \Accuracy{}  & 0.32     & 0.60          & 0.69          & 0.66          & 0.66          & \textbf{0.74} & 0.66          & 0.68          \\ \cline{2-10} 
                                 & \Precision{} & 0.24     & 0.60          & 0.66          & 0.65          & 0.65          & \textbf{0.73} & 0.65          & 0.67          \\ \cline{2-10} 
                                 & \Recall{}    & 0.25     & 0.60          & 0.69          & 0.66          & 0.66          & \textbf{0.74} & 0.66          & 0.68          \\ \cline{2-10} 
                                 & \FScore{}  & 0.23     & 0.57          & 0.65          & 0.63          & 0.64          & \textbf{0.72} & 0.64          & 0.66          \\ \hline
    \multirow{4}{*}{\GCJCPP}     & \Accuracy{}  & 0.90     & 0.98          & 0.99          & 0.94          & 0.98          & \textbf{1.00} & 0.99          & \textbf{1.00} \\ \cline{2-10} 
                                 & \Precision{} & 0.86     & 0.98          & 0.99          & 0.92          & 0.97          & \textbf{1.00} & 0.98          & \textbf{1.00} \\ \cline{2-10} 
                                 & \Recall{}    & 0.90     & 0.98          & 0.99          & 0.94          & 0.98          & \textbf{1.00} & 0.99          & \textbf{1.00} \\ \cline{2-10} 
                                 & \FScore{}  & 0.87     & 0.98          & 0.99          & 0.92          & 0.97          & \textbf{1.00} & 0.98          & \textbf{1.00} \\ \hline
    \multirow{4}{*}{\GCJJava{}}    & \Accuracy{}  & 0.94     & 0.98          & 0.98          & 0.98          & 0.98          & \textbf{0.99} & 0.99          & \textbf{0.99} \\ \cline{2-10} 
                                 & \Precision{} & 0.94     & \textbf{0.99} & \textbf{0.99} & \textbf{0.99} & \textbf{0.99} & \textbf{0.99} & \textbf{0.99} & \textbf{0.99} \\ \cline{2-10} 
                                 & \Recall{}    & 0.93     & 0.98          & 0.98          & 0.98          & 0.98          & \textbf{0.99} & \textbf{0.99} & \textbf{0.99} \\ \cline{2-10} 
                                 & \FScore{}  & 0.93     & 0.98          & 0.98          & 0.98          & 0.98          & \textbf{0.99} & 0.98          & \textbf{0.99} \\ \hline
    \multirow{4}{*}{\GCJPython{}}  & \Accuracy{}  & 0.71     & 0.92          & 0.93          & 0.91          & 0.92          & \textbf{0.98} & 0.94          & 0.96          \\ \cline{2-10} 
                                 & \Precision{} & 0.63     & 0.90          & 0.91          & 0.90          & 0.90          & \textbf{0.97} & 0.92          & 0.95          \\ \cline{2-10} 
                                 & \Recall{}    & 0.71     & 0.92          & 0.93          & 0.91          & 0.92          & \textbf{0.98} & 0.94          & 0.96          \\ \cline{2-10} 
                                 & \FScore{}  & 0.66     & 0.91          & 0.92          & 0.90          & 0.90          & \textbf{0.97} & 0.93          & 0.96          \\ \hline
    \multirow{4}{*}{\GithubC{}}    & \Accuracy{}  & 0.77     & 0.94          & \textbf{0.96} & 0.94          & 0.95          & \textbf{0.96} & 0.95          & 0.95          \\ \cline{2-10} 
                                 & \Precision{} & 0.71     & 0.94          & 0.96          & 0.94          & 0.95          & \textbf{0.97} & 0.95          & 0.96          \\ \cline{2-10} 
                                 & \Recall{}    & 0.71     & 0.94          & \textbf{0.96} & 0.94          & 0.95          & \textbf{0.96} & 0.95          & 0.95          \\ \cline{2-10} 
                                 & \FScore{}  & 0.69     & 0.93          & \textbf{0.96} & 0.93          & 0.94          & \textbf{0.96} & 0.94          & 0.95          \\ \hline
    \multirow{4}{*}{\GithubJava{}} & \Accuracy{}  & 0.97     & 0.98          & \textbf{1.00} & 0.99          & 0.98          & \textbf{1.00} & 0.99          & \textbf{1.00} \\ \cline{2-10} 
                                 & \Precision{} & 0.95     & 0.98          & \textbf{1.00} & 0.99          & 0.99          & \textbf{1.00} & 0.99          & \textbf{1.00} \\ \cline{2-10} 
                                 & \Recall{}    & 0.95     & 0.98          & \textbf{1.00} & 0.99          & 0.98          & \textbf{1.00} & 0.99          & \textbf{1.00} \\ \cline{2-10} 
                                 & \FScore{}  & 0.94     & 0.98          & 0.99          & 0.98          & 0.98          & \textbf{1.00} & 0.98          & \textbf{1.00} \\ \hline
    \end{tabular}}
    \end{table*}

\begin{table}[tb]
\centering
\caption{\Accuracy{} and \JSD{} of \PbNN{}}
\label{table:pbnn-accuracy-js-divergence}
\begin{tabular}{|c|cc|}
\hline
\multirow{2}{*}{\textbf{Dataset}} & \multicolumn{2}{c|}{\textbf{\PbNN{}}}                              \\ \cline{2-3}
                                  & \multicolumn{1}{c|}{\textbf{\Accuracy{}}} & \textbf{\JSD{}} \\ \hline
\GCJCPP{}                         & \multicolumn{1}{c|}{0.90}              & 0.55                   \\ \hline
\GCJJava{}                        & \multicolumn{1}{c|}{0.94}              & 0.71                   \\ \hline
\GCJPython{}                      & \multicolumn{1}{c|}{0.71}              & 0.43                   \\ \hline
\GithubJava{}                     & \multicolumn{1}{c|}{0.97}              & 0.92                   \\ \hline
\GithubC{}                        & \multicolumn{1}{c|}{0.77}              & 0.46                   \\ \hline
\LeetCode{}                       & \multicolumn{1}{c|}{0.32}              & 0.18                   \\ \hline
\end{tabular}
\end{table}

\revision{\emph{Performance differences across dataset sizes}:  Analysis of performance across datasets reveals that the LM advantage over \PbNN{} is not primarily driven by dataset size, but rather by dataset characteristics. Table~\ref{tab:caa_results} shows that on GitHub-Java (2,667 samples), \PbNN{} achieves 0.97 accuracy—competitive with LMs—while on the larger LeetCode dataset (4,753 samples), \PbNN{} accuracy drops to 0.32. This counterintuitive pattern is explained by samples-per-author density and author separability (Table~\ref{table:pbnn-accuracy-js-divergence}): GitHub-Java has high \JSD{} (0.92) with sufficient samples per author, enabling \PbNN{} to learn discriminative features. Conversely, LeetCode's 198 authors with sparse samples per author (average $\approx$24) result in low separability (\JSD{}: 0.18), where \PbNN{}'s limited capacity fails but \LLMsS{} leverage pre-trained representations to maintain 0.74 accuracy. Notably, even on the smallest dataset (GCJ-CPP, 160 samples), LMs achieve perfect accuracy while \PbNN{} reaches 0.90, demonstrating that pre-trained embeddings compensate for limited training data. The performance gap correlates strongly with task complexity (number of authors) and data sparsity per author, rather than absolute dataset size. \PbNN{} remains competitive only when: (1) high samples-per-author ratio enables sufficient feature learning, and (2) authors exhibit naturally high stylistic separability. }

\revision{\emph{Performance differences across languages}: Analysis across programming languages reveals that the LM advantage over \PbNN{} varies significantly by language characteristics. In language-specific datasets (Table~\ref{table:language-comparison}), \PbNN{} achieves competitive performance on Java datasets (GCJ-Java: 0.94, GitHub-Java: 0.97) but degrades on C/C++ (GitHub-C: 0.77, GCJ-CPP: 0.90) and Python (GCJ-Python: 0.71). This trend intensifies in the multi-lingual LeetCode dataset (Table~\ref{table:language-comparison}), where \PbNN{}'s accuracy ranges from 0.27 (C++) to 0.68 (Ruby)—a 41-point variance—while \DeepSeek{} maintains consistently high performance across all languages (0.72-0.95, only 23-point variance). The performance gap correlates with language complexity and stylistic diversity. \PbNN{} struggles most with syntactically flexible languages like C++ (gap: 0.45) and Java (gap: 0.43), where multiple idiomatic patterns dilute authorship signals that \PbNN{} can learn from limited training data. Conversely, the gap narrows for languages with stronger conventions like Ruby and C\#.}

\begin{table}[tb]
\centering
\caption{Language-Specific Performance: \PbNN{} vs Best LM}
\label{table:language-comparison}
\begin{tabular}{|l|c|c|c|}
\hline
\textbf{Language} & \textbf{\PbNN{}} & \textbf{Best LM} & \textbf{Gap} \\ \hline
\multicolumn{4}{|c|}{\textit{Monolingual Datasets}} \\ \hline
Java (GCJ)        & 0.94          & 0.99             & 0.05         \\ \hline
Java (GitHub)     & 0.97          & 1.00             & 0.03         \\ \hline
C++ (GCJ)         & 0.90          & 1.00             & 0.10         \\ \hline
C (GitHub)        & 0.77          & 0.96             & 0.19         \\ \hline
Python (GCJ)      & 0.71          & 0.98             & 0.27         \\ \hline
\multicolumn{4}{|c|}{\textit{LeetCode (Multilingual)}} \\ \hline
C++               & 0.27          & 0.72             & 0.45         \\ \hline
Java              & 0.32          & 0.75             & 0.43         \\ \hline
Python            & 0.48          & 0.77             & 0.29         \\ \hline
JavaScript        & 0.54          & 0.91             & 0.37         \\ \hline
C\#               & 0.67          & 0.95             & 0.28         \\ \hline
Ruby              & 0.68          & 0.79             & 0.11         \\ \hline
\end{tabular}
\end{table}

\revision{\emph{Performance differences across code lengths}: Code length analysis reveals contrasting dependencies between \PbNN{} and LMs. As shown in Table~\ref{table:length-analysis}, \PbNN{} exhibits erratic, non-monotonic performance across code lengths: starting at 0.22 accuracy for very short snippets ($\leq$50 tokens), peaking modestly at 0.35 for medium code (201--500 tokens), and then fluctuating between 0.29 and 0.33 for longer code (501--4000 tokens) without ever recovering its peak. This unpredictable pattern---with at most 0.06 absolute difference between its ``optimal'' bracket and any other bracket above 100 tokens---suggests \PbNN{} lacks robust length-invariant feature learning.}

\revision{In contrast, \DeepSeek{} (Best Performing LM) improves steadily with code length: 0.50 on short snippets ($\leq$50 tokens), rising to 0.82 for medium-long code (501--1000 tokens), and achieving perfect accuracy (1.00) on 2001--4000 token samples. The only departure from this trend is a marginal 0.02 dip at 1001--2000 tokens (0.80), well within fold-to-fold variation. The performance gap systematically widens with code length: a 0.28 gap for short code ($\leq$50 tokens) expanding to a 0.71 gap for long code (2001--4000 tokens). Even at \PbNN{}'s nominal peak (201--500 tokens: 0.35), \LLMsS{} achieve 0.76 accuracy---a 0.41 advantage. Most critically, \LLMsS{}' sustained improvement on longer code---where attribution is most valuable for identifying authors of complete functions, classes, or files---directly contradicts \PbNN{}'s plateau at approximately 0.35 accuracy. This demonstrates that pre-trained representations with superior positional encoding and attention mechanisms effectively capture authorship signals across all code lengths, while \PbNN{}'s learned embeddings fail to scale beyond superficial pattern matching.}

\begin{table}[tb]
\centering
\caption{Performance vs. Code Length: \PbNN{} and \DeepSeek{}}
\label{table:length-analysis}

\begin{tabular}{l|c|c|c}
\hline
\textbf{Token Length} & \textbf{\PbNN{}} & \textbf{\DeepSeek{}} & \textbf{Gap} \\ \hline
$\leq$50                   & 0.22          & 0.50              & 0.28         \\ \hline
51-100                & 0.26          & 0.63              & 0.37         \\ \hline
101-200               & 0.30          & 0.73              & 0.43         \\ \hline
201-500               & 0.35          & 0.76              & 0.41         \\ \hline
501-1000              & 0.31          & 0.82              & 0.51         \\ \hline
1001-2000             & 0.33          & 0.80              & 0.47         \\ \hline
2001-4000             & 0.29          & 1.00              & 0.71         \\ \hline
\end{tabular}
\end{table}

 \textbf{\LLMsS{} vs \LLMsS{}}: While \LLMsS{} consistently outperform \PbNN{}, their effectiveness varies depending on the dataset. Our analysis indicates that in some datasets, \LLMsS{} perform comparably with no statistically significant differences, whereas in others, certain models achieve notably better results. 

In \GCJCPP{} and \GCJJava{}, all \LLMsS{} except one achieved near-perfect scores with no statistically significant differences in performance. In \GCJPython{}, however, \UnixCoder{} and \DeepSeek{} significantly outperformed other models, with \DeepSeek{} achieving the highest \Accuracy{}, \Precision{}, \Recall{}, and \FScore{}. Other \LLMsS{} performed similarly in this dataset, showing no statistically significant differences. A similar trend is observed in \GithubC{}, where \DeepSeek{} and \CodeLlama{} outperform other models, with \DeepSeek{} achieving the highest scores across all metrics. In contrast, the remaining \LLMsS{} exhibit comparable performance. In \GithubJava{}, \UnixCoder{}, \DeepSeek{}, and \CodeLlama{} significantly outperform other models. As previously discussed, these datasets share a similar distribution of code samples and authors, which explains the limited performance variation among \LLMsS{}. In such cases, capacity factors like model number of parameters, and maximum input length appear to have a minimal impact on performance differences, as all top-performing models can effectively leverage the dataset's characteristics.

While the capacity factors of \LLMsS{} (such as the number of parameters and maximum input length) have shown minimal impact on performance differences across most datasets, the \LeetCode{} dataset presents a different scenario due to its multilingual nature, class imbalance, short code snippets, and minimal stylometric variation. Models with a smaller number of parameters (\smallLLMParameters{}) and similar input length (512 tokens), such as \CodeBERT{}, \ContraC{}, \ContraG{}, and \GraphCodeBERT{}, perform comparably. However, \UnixCoder{} (\UnixCoderParameters{}, 1024 tokens) and \DeepSeek{} (\DeepSeekParameters{}, 2500 tokens) significantly outperform the others, with \DeepSeek{} achieving the highest \Accuracy{}, \Precision{}, \Recall{}, and \FScore{}. This suggests that in datasets with these characteristics, \LLMsS{} with larger capacity are more effective for \CAAS{}.

Interestingly, although \CodeLlama{} has the largest model size and input length, it performs comparably to \UnixCoder{} and \DeepSeek{}. This is likely because its parameters are not fully fine-tuned for the \CAAS{} task. Thus, we conclude that larger \LLMsS{} are more effective for \CAAS{} in multilingual, imbalanced datasets with short code snippets and minimal stylometric variation, provided that they are appropriately fine-tuned.

\begin{table}[tb]
    \setlength{\tabcolsep}{2pt}
    \caption{Language-wise \Accuracy{} of the \LLMsS{} in \LeetCode{}. Samples per language: 62.19\% \cpp{}, 24.83\% \java{}, 6.77\% \python{}, 3.79\% \javascript{}, 2.02\% \csharp{}, 0.40\% \ruby{}.}
    \label{tab:language-wise-results}
    \centering
    \scalebox{1}{
    \begin{tabular}{l|c|c|c|c|c|c}
    \hline
    Model         & \cpp{}  & \java{} & \python{} & \javascript{} & \csharp{} & \ruby{} \\ \hline
    \CodeBERT{}      & 0.55 & 0.58 & 0.74   & 0.82       & 0.93   & 0.79 \\ \hline
    \ContraC{} & 0.63 & 0.62 & 0.72   & 0.86       & 0.94   & 0.74 \\ \hline
    \ContraG{} & 0.62 & 0.63 & 0.73   & 0.79       & 0.91   & 0.74 \\ \hline
    \GraphCodeBERT{} & 0.63 & 0.62 & 0.74   & 0.80       & 0.95   & 0.68 \\ \hline
    \UnixCoder{}     & 0.66 & 0.66 & 0.78   & 0.88       & 0.94   & 0.89 \\ \hline
    \DeepSeek{}      & 0.72 & 0.75 & 0.77   & 0.91       & 0.95   & 0.79 \\ \hline
    \CodeLlama{} & 0.67 & 0.66 & 0.68   & 0.79       & 0.95   & 0.79 \\ \hline
    Average       & 0.64 & 0.65 & 0.74   & 0.84       & 0.94   & 0.77 \\ \hline
    \end{tabular}}
    \end{table}

 \textbf{Generalization}: A critical aspect of evaluating \LLMsS{} for \CAAS{} is understanding how well their performance generalizes across different dataset characteristics, including monolingual vs. multilingual code, balanced vs. imbalanced distributions, and variations in the length of the code snippet and stylometric diversity. Our analysis reveals that while \LLMsS{} show strong generalization in well-balanced monolingual datasets, their effectiveness varies significantly in more complex multilingual and imbalanced settings.

For example, in well-balanced monolingual datasets such as \GCJCPP{}, \GCJJava{}, \GCJPython{}, \GithubC{}, and \GithubJava{}, \LLMsS{} achieve an average \Accuracy{} of 96\%, indicating strong generalization across these datasets. However, the \LeetCode{} dataset presents additional challenges, such as short code snippets and reduced stylometric variation. As shown in Table~\ref{tab:caa_results}, the average \Accuracy{} drops to 67\%---and reaches only 74\% for the best-performing model---suggesting that \LLMsS{} struggle to generalize as effectively in this setting. Table~\ref{tab:language-wise-results} breaks this result down by language.

\begin{figure}[tb]
    \centering
    \subfloat[\CodeBERT{}'s Accuracy by Code Length]{
        \includegraphics[width=0.48\textwidth]{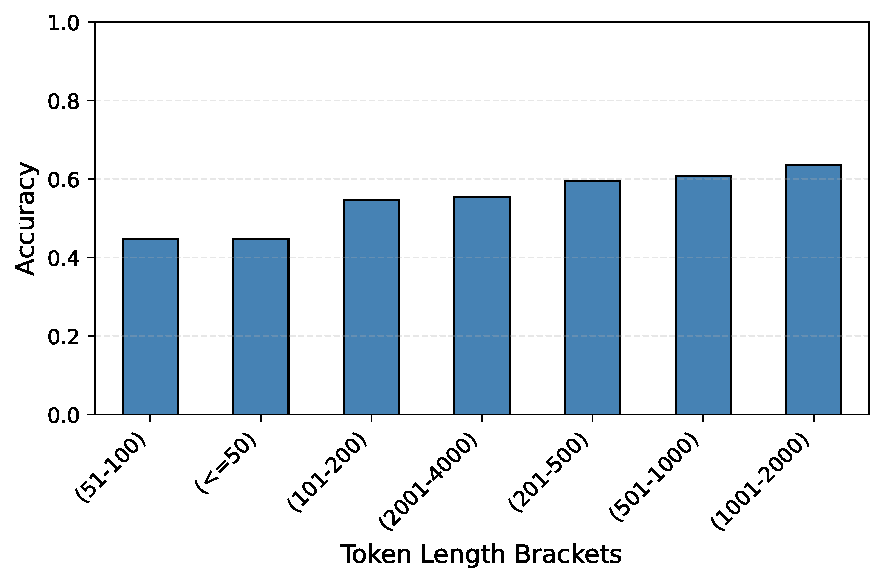}
        \label{fig:codebert-length}
    }
    \hfill
    \subfloat[\DeepSeek{}'s Accuracy by Code Length]{
        \includegraphics[width=0.48\textwidth]{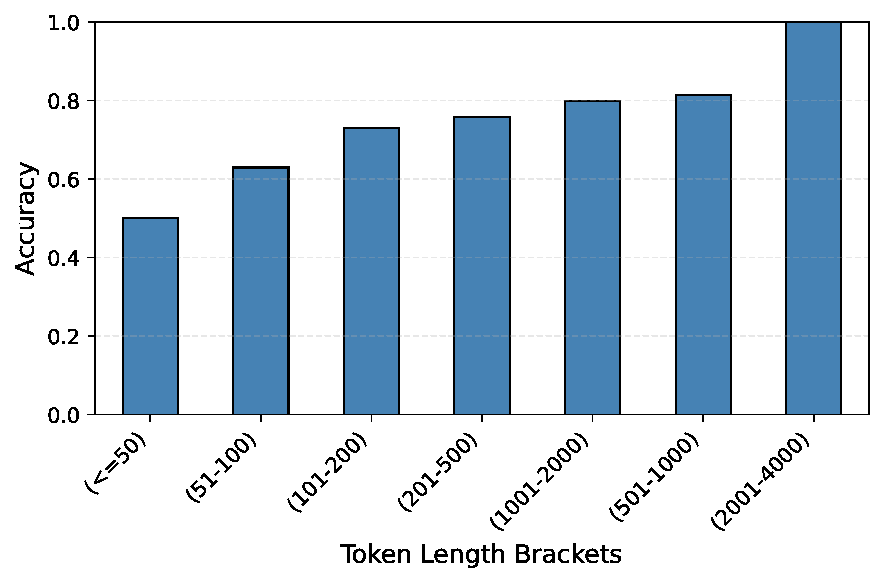}
        \label{fig:deepseek-length}
    }
    \caption{\revision{Performance analysis across different length brackets of different models}}
    \Description{Two line charts showing accuracy versus code length brackets for \CodeBERT{} and \DeepSeek{}, demonstrating that accuracy increases with longer code snippets.}
    \label{fig:length-brackets-performance}
\end{figure}

\revision{Models tend to perform better with longer code snippets, as the mean and median lengths of incorrectly attributed samples are notably shorter than those correctly attributed. We analyzed model performance across seven token length brackets: $\leq$50, 51--100, 101--200, 201--500, 501--1000, 1001--2000, and 2001--4000 (Figure~\ref{fig:length-brackets-performance}). All models show improved accuracy with longer code, but patterns differ by architecture. As Figure~\ref{fig:codebert-length} illustrates, BERT-based models (\CodeBERT{}, \GraphCodeBERT{}, \UnixCoder{}, \ContraC{}, and \ContraG{}) demonstrate relatively stable performance across lengths, with 10--20 percentage point improvements from the shortest to the longest brackets. As Figure~\ref{fig:deepseek-length} shows, larger models (\CodeLlama{}, \DeepSeek{}) exhibit greater length sensitivity, with gains of 20--30 percentage points or more, stabilizing around 200--500 tokens. These findings show that code snippets should contain at least 200 tokens for reliable attribution using larger models.}

Another key observation from Table~\ref{tab:language-wise-results} is that having more training data does not necessarily improve \Accuracy{}. For example, \cpp{} accounts for the largest proportion of samples (62.19\%) in the \LeetCode{} dataset, yet the average \Accuracy{} of \LLMsS{} is only 64\%. In contrast, with just 2.02\% of the dataset, \csharp{} achieves a significantly higher \Accuracy{} of 94\%. The reason for this discrepancy is that when more developers solve a problem using the same language, the solutions exhibit \textit{less stylometric variation}, making authorship attribution more difficult.

For example, the \cpp{} solutions to the \textit{Power of Two} problem~\cite{LeetCodePowerOfTwo} are among the most misattributed samples in our results. This problem has two main solutions with different time complexities: \( O(\log N) \) and \( O(1) \). Since the \( O(\log N) \) solution is less efficient, most users opt for the \( O(1) \) approach, leading to highly similar solutions. As a result, the uniformity in coding patterns reduces distinguishable stylometric differences, making it harder for \LLMsS{} to correctly attribute authorship. \revision{We also measure \JSD{} in \cpp{} and \csharp{} samples to quantify stylometric variation. Our results show that the \JSD{} of \cpp{} samples is lower than that of \csharp{} samples: as \JSD{} increases from 0.26 (\cpp{}) to 0.58 (\csharp{}), accuracy increases from 0.72 to 0.95.}

A correlation analysis between the number of samples per author and \Accuracy{} reveals a weak but statistically significant correlation (0.17). Although it is generally expected that authors with more samples would have higher attribution accuracy, the lack of strong correlation suggests that \textit{having more samples does not necessarily aid authorship attribution when stylometric variation is low}.

Taking everything together, these findings highlight that while \LLMsS{} generalize well in larger, balanced datasets, their performance deteriorates for shorter code snippets that contain less stylometric variation, underscoring the need for further improvements in handling such complexities.

\textbf{Computational efficiency}: \revision{BERT-based models (\CodeBERT{}, \GraphCodeBERT{}, \ContraC{}, \ContraG{}, \UnixCoder{}) achieve inference times of 0.0015--0.0032 seconds with less than 1.5\,GB of memory, making them suitable for real-time applications. In contrast, \CodeLlama{} (7B) requires 30.2\,GB of VRAM and 0.167 seconds per prediction---approximately 100$\times$ slower---while \DeepSeek{} (1.3B) offers a middle ground at 21.5\,GB and 0.059 seconds.} \revision{For practitioners, this suggests that BERT-based models are preferable for resource-constrained or latency-sensitive applications (e.g., bug triaging, CI/CD pipelines), while larger \LLMsS{} may be justified for offline analysis where accuracy is paramount.}

\finding{This research question evaluates whether language models (LMs) can effectively attribute code snippets to their correct authors across a range of realistic conditions. The study benchmarks seven transformer-based LMs on six diverse datasets and compares them to the prior state-of-the-art model \PbNN{}. Results show that LMs consistently outperform \PbNN{}, with large models like \DeepSeek{} and \CodeLlama{} excelling on multilingual, imbalanced, and short-code datasets like LeetCode. Smaller LMs such as \CodeBERT{} perform well on balanced, monolingual datasets. Notably, code stylometric variation and snippet length impact attribution more than the volume of training samples, suggesting that LMs are sensitive to the nuanced structure of coding styles.}

\subsection{\RQ{2}: \LLMsS{}' Code Stylometry Understanding}
\textbf{Qualitative}: To determine whether \LLMsS{} truly learn orthogonal coding styles, it is crucial to first assess the faithfulness of the interpretability method used — \xaiMethod{}. Faithfulness ensures that the feature attributions provided by the method accurately reflect the model's true decision-making process. Without faithful explanations, conclusions about orthogonal coding styles may be misleading.

\begin{figure}
    \centering
    \includegraphics[width=0.7\textwidth]{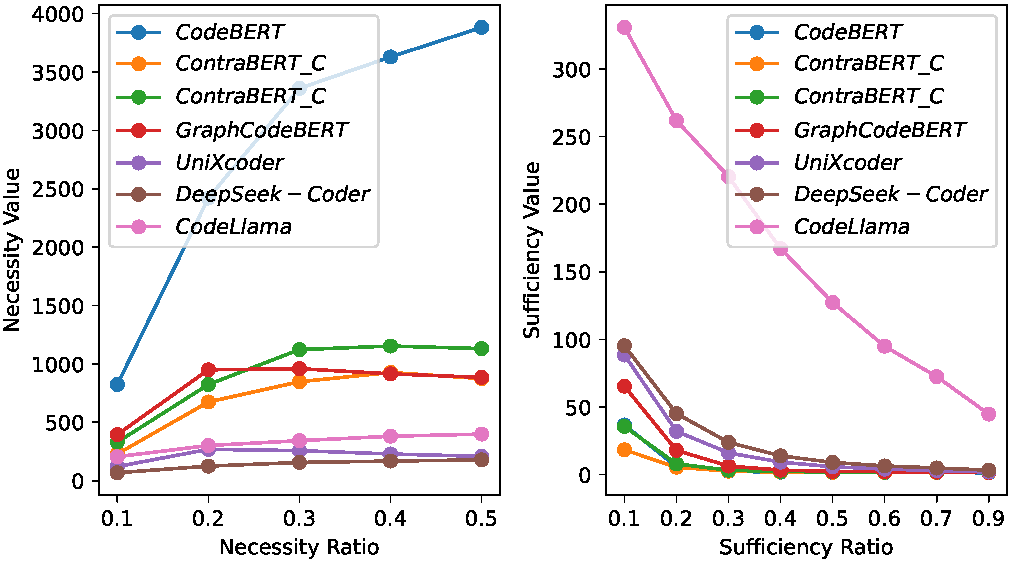}
    \caption{Faithfulness of \xaiMethod{}'s explanations}
    \Description{Line charts showing necessity and sufficiency scores for \xaiMethod{} explanations across language models at varying feature removal percentages.}
    \label{fig:necessity_sufficiency}
\end{figure}

To ensure faithful attribution of model predictions, we apply \xaiMethod{} to the embedding layers of \LLMsS{}. The embedding layer serves as the model's initial processing stage, where raw code tokens are transformed into numerical representations. By computing feature attributions at this level, we ensure that the importance scores correspond directly to input tokens rather than higher-level latent representations. As shown in Figure~\ref{fig:necessity_sufficiency}, \Necessity{} values are consistently high across most \LLMsS{} up to 30\%, indicating that removing the 30\% most important features identified by \xaiMethod{} significantly reduces model confidence. This suggests that \xaiMethod{} correctly identifies the 30\% of features that are critical for classification. Additionally, \Sufficiency{} values also consistently decrease up to 30\%, meaning that the \LLMsS{} can make confident predictions when retaining only the 30\% of features identified as important. Therefore, we use the top 30\% most important features for our analysis of orthogonal coding styles, as this provides a balance between capturing critical information and interpretability.

\begin{figure}
    \centering
    \includegraphics[width=0.4\columnwidth]{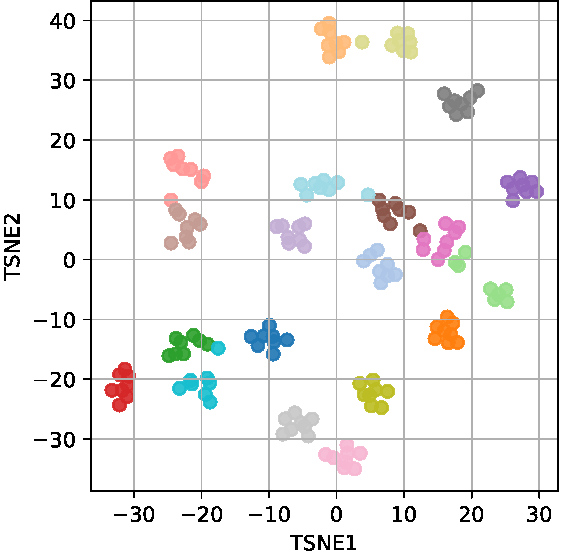}
    \caption{Visualization of \DeepSeek{}'s explanations}
    \Description{t-SNE scatter plot showing clusters of code samples from different authors based on DeepSeek-Coder's top feature attributions, with distinct author clusters visible.}
    \label{fig:tsne_plot}
\end{figure}

Before analyzing the orthogonal coding styles of different authors, we first examine whether \LLMsS{} capture consistent coding styles among the samples of the same author. Suppose a particular \LLM{} correctly attributes multiple code snippets ($C_1$, $C_2$, $C_3$, ... $C_n$) to author $X_1$. If the model truly relies on the top 30\% most important features identified by \xaiMethod{}, we should observe a significant overlap in the important features across these samples. Our analysis reveals that the top 30\% important features of $C_1$, $C_2$, $C_3$, ... $C_n$ exhibit a high degree of overlap (ranging from 84\% to 94\%), indicating that \LLMsS{} capture consistent coding styles for the same author. Additionally, Figure~\ref{fig:tsne_plot} presents a t-SNE~\cite{vanDerMaaten2008} visualization of the top 30\% important features extracted from code samples in the \GCJCPP{} dataset, where \xaiMethod{} is applied to \DeepSeek{}. The figure shows that samples from the same author form distinct clusters, further confirming that \DeepSeek{} consistently relies on similar stylistic features within an author's code samples. We also observe similar results for other \LLMsS{}.

\begin{figure}
    \centering
    \includegraphics[width=0.7\columnwidth]{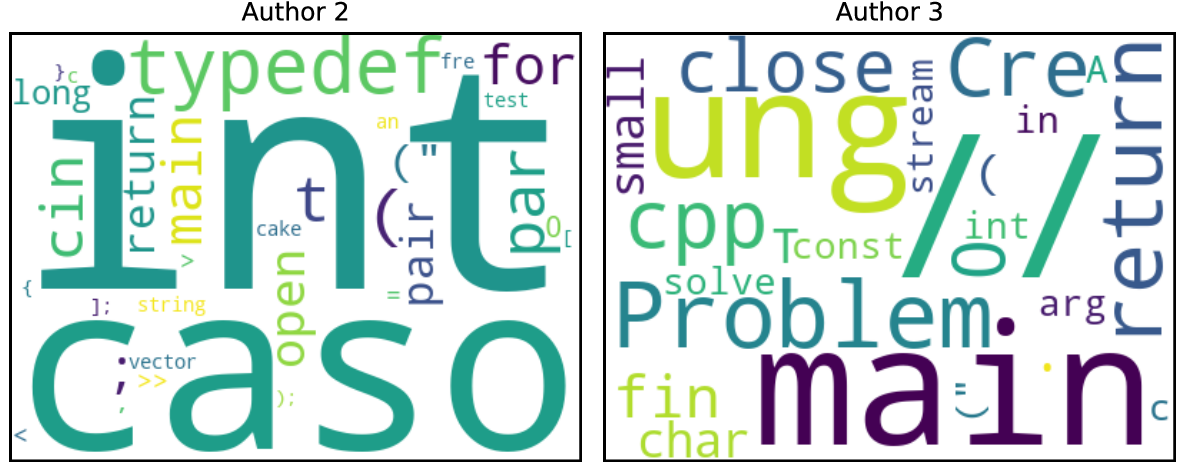}
    \caption{Visualization of orthogonal coding styles across different authors using word clouds.} 
    \Description{Word clouds for different authors showing their most influential code tokens identified by \xaiMethod{}, highlighting distinct coding style preferences.}
    \label{fig:orthogonal_stylometry}
\end{figure}

From Figure~\ref{fig:tsne_plot}, we observe that code samples from different authors are well separated, indicating that \LLMsS{} capture distinct, orthogonal coding styles across authors. To further investigate this, we visualize the top 30\% important features of different authors using word clouds. The word clouds in Figure~\ref{fig:orthogonal_stylometry} illustrate that different authors exhibit unique stylistic preferences, as evidenced by the dominant keywords and patterns in their respective clouds. For example, Author 2 frequently uses the keyword `int', suggesting a preference for integer-based operations, while Author 3 consistently includes `//' comments, indicating a habit of documenting their code. This reinforces the idea that \LLMsS{} effectively capture orthogonal coding styles by recognizing individualized patterns in variable usage, commenting habits, and other stylistic features.

Somewhat interestingly, we observe that different \LLMsS{} capture distinct orthogonal coding styles for the same author. This variation arises due to differences in tokenization, as each \LLM{} employs a unique tokenizer that affects how the same code snippet is segmented into tokens. For example, \DeepSeek{} treats `addTwoNumbers' as four tokens (`add', `Two', `Num', `bers'), whereas \CodeLlama{} splits it into three tokens (`\_add', `Two', `Numbers'). Consequently, \xaiMethod{} identifies different important features across models, leading to variation in feature attributions and contributing to the orthogonal behavior observed in \LLMsS{}.

\begin{figure}
    \centering
    \includegraphics[width=0.5\columnwidth]{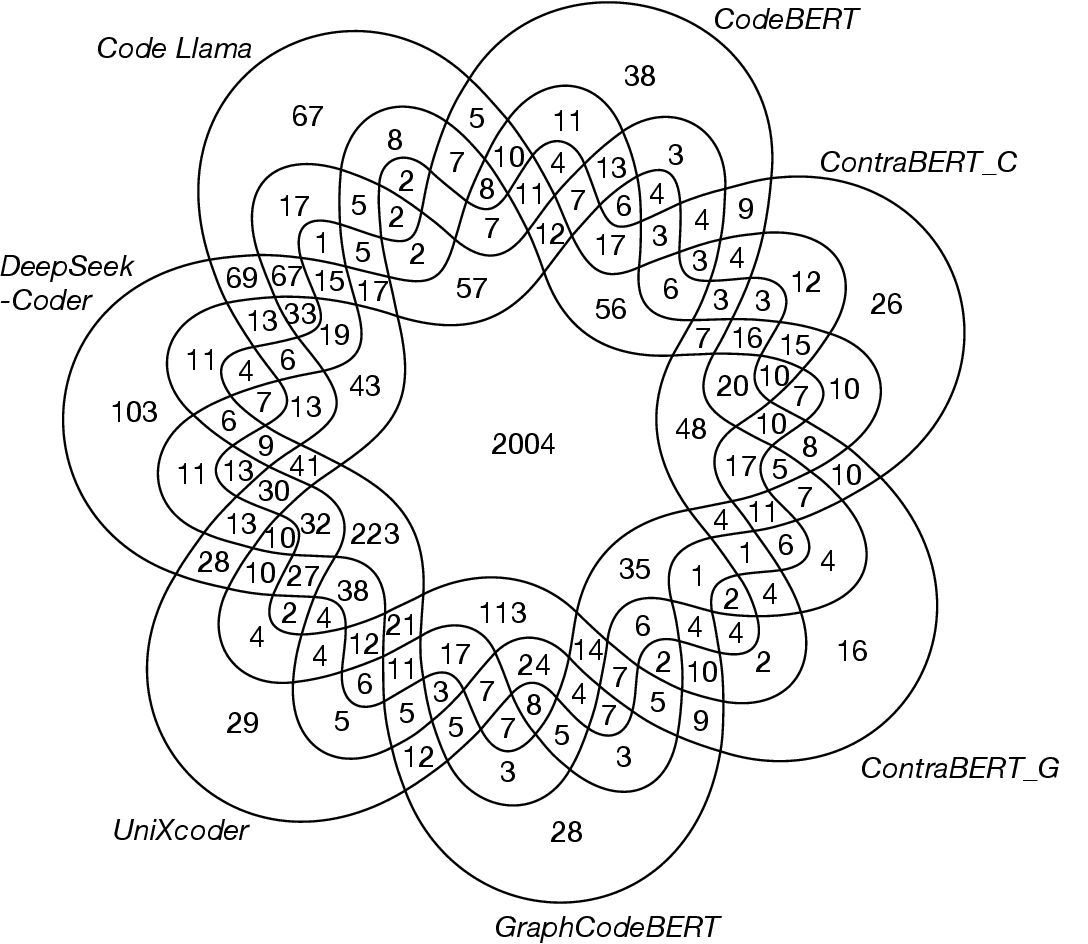}
    \caption{Orthogonality of \LLMsS{} in \LeetCode{} dataset}
    \Description{Venn diagram showing the degree of orthogonality among language models, with shared and exclusive correctly attributed code samples in the LeetCode dataset.}
    \label{fig:llm_Orthogonality}
\end{figure}

Figure~\ref{fig:llm_Orthogonality} illustrates the Degree of Orthogonality~\cite{dipongkor2023comparative} in \CAAS{} within the \LeetCode{} dataset. The value in the common area represents the number of code snippets correctly attributed by all \LLMsS{}, while the exclusive regions indicate the number of samples uniquely classified by each model, highlighting their individual strengths. For instance, all \LLMsS{} correctly assign 2,004 code snippets to the original developer in this dataset. However, \DeepSeek{} exhibits the highest Degree of Orthogonality in \CAAS{}, followed by \CodeLlama{} and \CodeBERT{}.

\begin{figure}[tb]
    \centering
    \subfloat[\DeepSeek{}'s Similarity Distribution]{
        \includegraphics[width=0.45\textwidth]{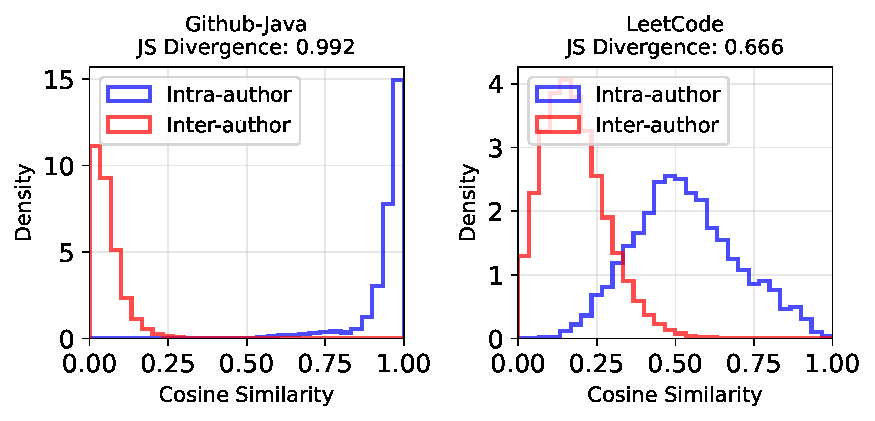}
        \label{fig:deep-intra-inter-sim-dist}
    }
    \hfill
    \subfloat[\PbNN{}'s Similarity Distribution]{
        \includegraphics[width=0.45\textwidth]{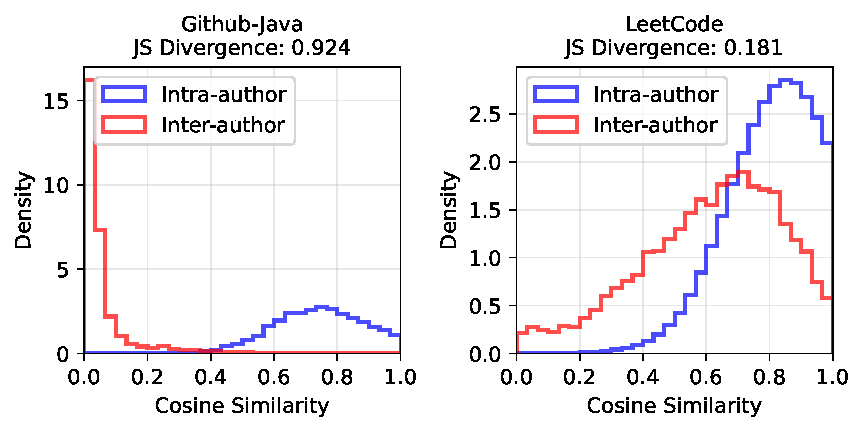}
        \label{fig:pbnn-intra-inter-sim-dist}
    }
    \caption{\revision{Intra-author vs. Inter-author Similarity Distributions of \DeepSeek{} and \PbNN{} Across Datasets}}
    \Description{Two panels of density plots comparing intra-author and inter-author cosine similarity distributions for DeepSeek-Coder and PbNN across multiple datasets.}
    \label{fig:intra-inter-sim-dist}
\end{figure}

\textbf{Quantitative}: \revision{Figure~\ref{fig:deep-intra-inter-sim-dist} presents the distributional analysis of intra-author versus inter-author cosine similarities in the \GithubJava{} and \LeetCode{} datasets using fine-tuned \DeepSeek{} models, while Figure~\ref{fig:pbnn-intra-inter-sim-dist} presents the corresponding distributions for \PbNN{}. The \JSD{} values quantify feature orthogonality and show strong correlation with classification performance (Table~\ref{tab:caa_results}).}

\minorrevision{Datasets with high classification accuracy generally exhibit correspondingly high \JSD{} values. \GithubJava{} achieves 100\% accuracy with a \JSD{} of 0.99, indicating strong distributional separation between intra-author and inter-author similarities. The intra-author distribution concentrates at high similarity values with minimal overlap with the inter-author distribution. Similarly, \GCJPython{} and \GithubC{} show strong separation with high performance. Conversely, the \LeetCode{} dataset demonstrates moderate separation (\JSD{}: 0.67) with 74\% accuracy. The 198-author diversity and varied coding contexts result in greater distributional overlap, making author distinction more difficult.}

\minorrevision{\JSD{} is generally predictive of classification accuracy across datasets with sufficient test set size. \GCJCPP{} appears as an exception (\JSD{}: 0.66, 100\% accuracy), but this result should be interpreted with caution: 8-fold CV over only 160 samples yields approximately 20 test samples per fold, which is too small to serve as a reliable indicator of true generalization performance. Excluding this low-confidence data point, the correlation between \JSD{} and accuracy holds consistently across the remaining datasets. This validates that \LLMsS{} learn genuine stylometric features rather than spurious correlations. Even in the most challenging scenario (\LeetCode{}), the \JSD{} of 0.67 indicates substantial distributional separation, confirming that models consistently capture meaningful, author-specific features.}

\finding{This RQ investigates whether LMs genuinely learn distinct, author-specific stylistic features or rely on spurious correlations. We apply \xaiMethod{} to derive feature attributions and measure explanation faithfulness using necessity and sufficiency metrics. Results demonstrate that top features contribute significantly to predictions, indicating faithful explanations. t-SNE visualizations and word clouds confirm that LMs consistently recognize author-specific coding habits and produce orthogonal representations across authors. Moreover, different LMs often rely on different stylistic cues, highlighting their complementary behaviors and opening avenues for ensemble or knowledge distillation strategies.}

\subsection{\RQ{3}: \LLMsS{}' Robustness to Adversarial Attacks}

\textbf{\GPT{}-based attack}: Since we used \GPT{} to generate adversarial samples, we first evaluate its effectiveness in generating these samples before assessing the robustness of \LLMsS{} against adversarial attacks. To measure robustness, we use the \textbf{adversarial success rate}~\cite{Ropgen}. Here, a higher adversarial success rate indicates that an \LLM{} is \textbf{less robust} against adversarial attacks.

In total, \allModelCorrectAttributions{} samples from the \LeetCode{} dataset were correctly attributed by at least one \LLM{}. Each of these samples was transformed into an adversarial version using \GPT{} with a predefined transformation prompt. Since we employed 28 different transformation prompts, the total number of adversarial samples generated was 114,016 (\allModelCorrectAttributions{} $\times$ 28). To ensure that the transformations preserved the original code's logic, we submitted all generated samples to \LeetCode{}. However, only 41.85\% (47,716) of the samples were accepted, meaning they retained functional correctness.

Next, we applied a three-step verification process using our parser to ensure that only valid adversarial samples were used for evaluation. Only 35,498 samples passed all three verification criteria and were subsequently used to evaluate the robustness of \LLMsS{} against adversarial attacks.

\begin{enumerate}
    \item \textbf{Applicability Check}: Was the requested transformation applicable to the original sample? (\textbf{Expected: YES}). For example, if the prompt instructed \GPT{} to remove all comments, but the original code contained no comments, the transformation would be invalid.
    \item \textbf{Correctness Check}: Was the requested transformation accurately applied to the adversarial code? (\textbf{Expected: YES}). For example, if \GPT{} was instructed to remove comments, the adversarial code should not contain comments.
    \item \textbf{Consistency Check}: Did \GPT{} introduce any unintended transformations? (\textbf{Expected: NO}). For example, if \GPT{} was instructed to replace a `for' loop with a `while' loop but also removed comments, this would introduce an unintended modification, making it difficult to isolate the cause of attack success.
\end{enumerate}

\begin{figure}
    \centering
    \includegraphics[width=0.7\columnwidth]{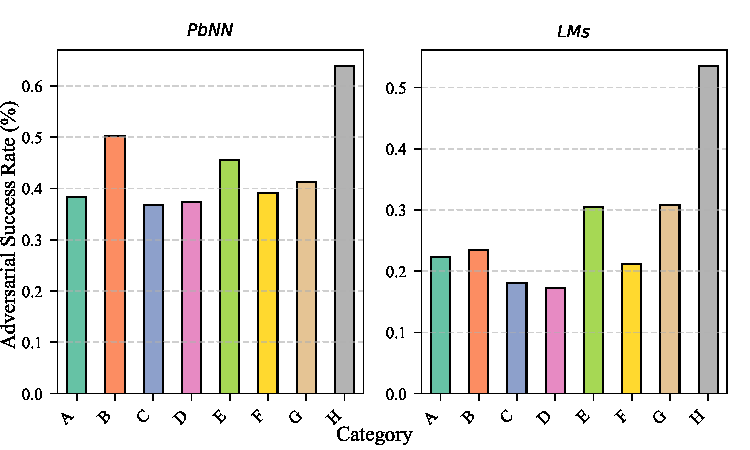}
    \caption{Adversarial Success Rate of \LLMsS{}}
    \Description{Grouped bar chart showing adversarial success rates for each transformation category across PbNN and seven language models.}
    \label{fig:adversarial_success_rate}
\end{figure}

 \textbf{SOTA vs \LLMsS{}}:  
Figure~\ref{fig:adversarial_success_rate} illustrates the adversarial success rate of \PbNN{} and various \LLMsS{}. The results indicate that \PbNN{} has a high adversarial success rate (\PbNNAdversarialSuccessRate{}), making it more susceptible to adversarial attacks than every \LLM{} we study except \CodeLlama{} (\CodeLlamaAdversarialSuccessRate{}), which we were only able to partially fine-tune via LoRA. Additionally, \PbNN{}'s category-wise success rate is the highest across all categories.

\textbf{\LLMsS{} vs \LLMsS{}}:  
Among the \LLMsS{}, \CodeLlama{} exhibits the highest adversarial success rate (\CodeLlamaAdversarialSuccessRate{}), followed by \DeepSeek{} (\DeepSeekAdversarialSuccessRate{}), \UnixCoder{} (\UnixCoderAdversarialSuccessRate{}), \GraphCodeBERT{} (\GraphCodeBERTAdversarialSuccessRate{}), \CodeBERT{} (\CodeBERTAdversarialSuccessRate{}), \ContraG{} (\ContraGAdversarialSuccessRate{}), and \ContraC{} (\ContraCAdversarialSuccessRate{}). Since we were unable to fine-tune all parameters of \CodeLlama{}, this may explain its comparatively lower robustness.  

Excluding \CodeLlama{}, Table~\ref{tab:code_models}, Table~\ref{tab:caa_results}, and Figure~\ref{fig:adversarial_success_rate} lead to the following key observations:  

\begin{enumerate}
    \item \textbf{More effective \LLMsS{} tend to be more vulnerable to adversarial attacks.}  
    For instance, \DeepSeek{}, which demonstrates the highest effectiveness in \CAAS{}, is also the most susceptible to adversarial attacks. However, one might argue that the total number of adversarial samples is not the same across all \LLMsS{}, making direct comparisons unfair. To address the sample imbalance concern, we re-conducted adversarial attacks using only the samples that were correctly attributed by all \LLMsS{}. The results remained consistent with our previous findings. For example, the adversarial success rates under equal sample conditions for \CodeBERT{}, \ContraC{}, \ContraG{}, \GraphCodeBERT{}, \UnixCoder{}, and \DeepSeek{} are \CodeBERTAdversarialSuccessRateEQ{}, \ContraCAdversarialSuccessRateEQ{}, \ContraGAdversarialSuccessRateEQ{}, \GraphCodeBERTAdversarialSuccessRateEQ{}, \UnixCoderAdversarialSuccessRateEQ{}, and \DeepSeekAdversarialSuccessRateEQ{}, respectively.

    \item \textbf{Smaller \LLMsS{} exhibit greater robustness to adversarial attacks.}  
    Models with fewer parameters, such as \CodeBERT{}, \ContraC{}, \ContraG{}, and \GraphCodeBERT{} (each having \smallLLMParameters{} parameters), show lower adversarial success rates than larger models like \UnixCoder{} (\UnixCoderParameters{}) and \DeepSeek{} (\DeepSeekParameters{}).

\end{enumerate}

Figure~\ref{fig:adversarial_success_rate} also presents the success rates of individual transformation rules for all \LLMsS{}. The results indicate that the following transformations are the most effective in deceiving \LLMsS{}.  

\begin{itemize}
    \item \revision{\textbf{E. Loop Transformations}:} \revision{Changing `for' statements to `while' statements \& vice versa. Figure~\ref{fig:adversarial-attack} shows an example of this transformation.}
    \item \revision{\textbf{G. Function Transformations}:} \revision{Swapping the order of function parameters, adding an extra integer parameter with a default value of zero, creating a new function from a group of statements, and swapping the order of function declarations. Figure~\ref{fig:category-h} shows an example of this transformation.}

    \item \revision{\textbf{H. Miscellaneous Transformations}:} \revision{Removing comments, unused code, and adding print/log statements to track variable values. Figure~\ref{fig:category-g} shows an example of this transformation.}
\end{itemize}

\begin{figure}[t]
    \centering
    \begin{minipage}[t]{0.48\columnwidth}
        \centering
        \includegraphics[width=\linewidth]{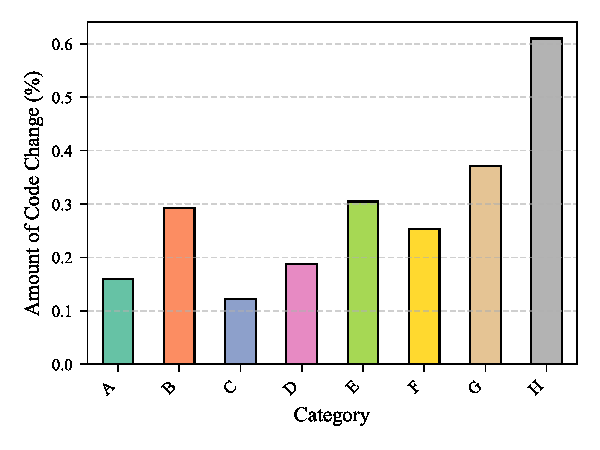}
        \caption*{(a) Extent of Code Changes per Rule.}
    \end{minipage}
    \hfill
    \begin{minipage}[t]{0.48\columnwidth}
        \centering
        \includegraphics[width=\linewidth]{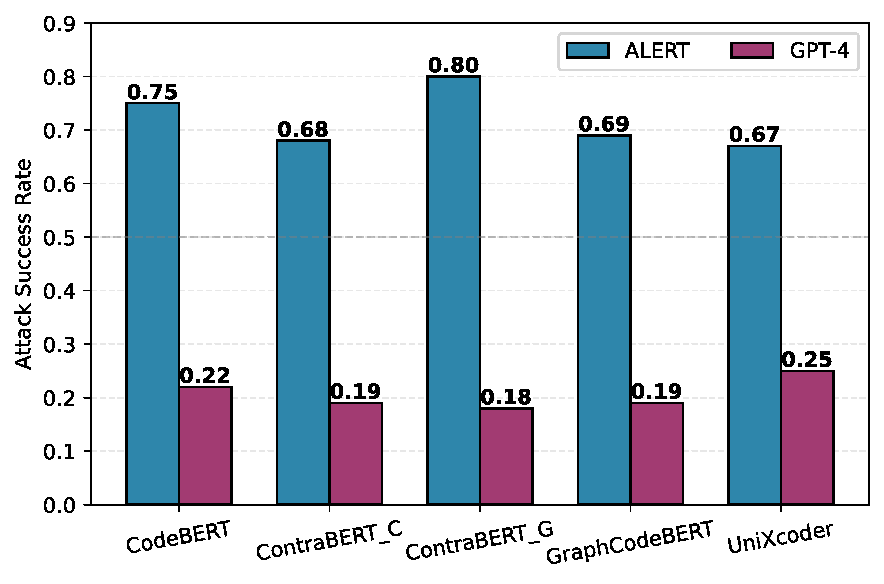}
        \caption*{(b) Attack Success Rates: ALERT vs. GPT-4-based Method.}
    \end{minipage}

    \caption{Comparison of code-change patterns and attack success rates across evaluation settings.}
    \Description{Two side-by-side bar charts: (a) showing the extent of code changes per adversarial transformation rule, and (b) comparing attack success rates between ALERT and GPT-4-based methods.}
    \label{fig:combined_results}
\end{figure}

To understand why these rules are particularly effective, we examined the amount of code they modify. As shown in Figure~\ref{fig:combined_results}-a, these transformation rules introduce more code changes than others. From Figure~\ref{fig:combined_results}-a, it is evident that \textit{transformation rule H} results in the highest number of code modifications and is also the most effective adversarial transformation in Figure~\ref{fig:adversarial_success_rate}. These findings suggest that transformation rules that modify more lines of code tend to be more effective in deceiving \LLMsS{}. The total amount of code changes is computed as:

\begin{equation}
    \frac{\text{\# of lines added, modified, and deleted in adversarial code}}{\text{Total \# of lines in original code}} \times 100\revision{\%}
\end{equation}

 \textbf{Why are \LLMsS{} less vulnerable to adversarial attacks?}  
Although adversarial success rates vary across \LLMsS{}, they are generally less susceptible to adversarial attacks compared to \PbNN{}. This is because \LLMsS{} rely on subtle syntactic and structural features of the code, such as keywords (e.g., \texttt{int}, \texttt{public}, \texttt{private}, \texttt{long}), braces (\{\}), and parentheses (()), as illustrated in Figure~\ref{fig:orthogonal_stylometry}. Modifying these elements while preserving semantics and successfully evading \LLMsS{} is inherently challenging, making them more robust against adversarial perturbations.

\revision{\textbf{ALERT-based attack}:} \revision{Figure~\ref{fig:combined_results}-b presents the attack success rates (ASR) of variable renaming attacks comparing ALERT and our GPT-4-based method across five pre-trained models of code: \CodeBERT{}, \ContraC{}, \ContraG{}, \GraphCodeBERT{}, and \UnixCoder{}. ALERT achieves substantially higher attack success rates compared to our GPT-4-based approach, with ASRs ranging from 0.67 to 0.80 (average: 0.72) versus 0.18 to 0.25 (average: 0.21) for GPT-4. The superior performance of ALERT can be attributed to its access to the victim models' internal representations. Specifically, ALERT leverages the masked language modeling (MLM) function and contextualized embeddings from the underlying pre-trained architectures to generate semantically natural variable substitutes. Furthermore, ALERT computes an Overall Importance Score (OIS) for each variable by measuring how its replacement affects the model's confidence in the ground truth label, allowing it to strategically prioritize which variables to replace first. This combination of insider knowledge about model internals and OIS-guided variable selection enables ALERT to craft adversarial examples that are specifically optimized to exploit the weaknesses of each model's learned representations.}

\revision{However, this advantage comes at a significant practical cost. First, ALERT's architecture-specific design limits its applicability—we could not apply ALERT to two additional models in our study (\DeepSeek{} and \CodeLlama{}) because they were not pre-trained with an MLM objective, which is fundamental to ALERT's substitute generation mechanism. This architectural dependency highlights a critical limitation: ALERT cannot generalize to models with different pre-training objectives. Second, in real-world code authorship attribution scenarios, such gray-box attacks are impractical because victim models are typically deployed remotely as black-box services~\cite{papernot2017practical,tramer2016stealing}. Attackers would only have query access to model predictions without any access to internal representations, model parameters, or the underlying pre-trained architecture. As noted by~\cite{quiring2019misleading}, attacks that rely on gradient information or internal model access have limited applicability in deployed systems where models are protected behind APIs. Our GPT-4-based method, despite achieving lower ASRs, represents a more realistic threat model as it operates under true black-box conditions—requiring only query access to model predictions. This makes our approach more practical and generalizable across different authorship attribution systems, regardless of their underlying architectures, pre-training objectives, or deployment configurations.}

\finding{This question assesses the robustness of LMs when exposed to adversarial code transformations that preserve functionality. Using 28 GPT-4-generated transformation prompts, we create over 114k samples and validate them using LeetCode submission and custom parsers. Results show that LMs are generally more robust than \PbNN{}, particularly because they learn deeper syntactic and structural patterns. Smaller LMs exhibit greater robustness, and contrastive learning (as in ContraBERT variants) improves resilience. The most effective attacks involve transformations that modify many tokens (e.g., control flow and comment removal), and overall, larger models tend to be more vulnerable unless carefully fine-tuned.}

\section{Discussion}
In this section, we discuss the important findings from all of our research questions.
\revision{\subsection{What are the common causes of misattribution?}}
\revision{We analyzed all misattributed samples across datasets and models, examining: (1) intra-author vs. inter-author similarity using cosine similarity of code embeddings, (2) stylometric variation using \JSD{}, and (3) code snippet length distribution. This systematic breakdown reveals the underlying causes of attribution failures. Primary error causes are:}

\begin{figure}[t]
\centering
\includegraphics[width=0.5\columnwidth]{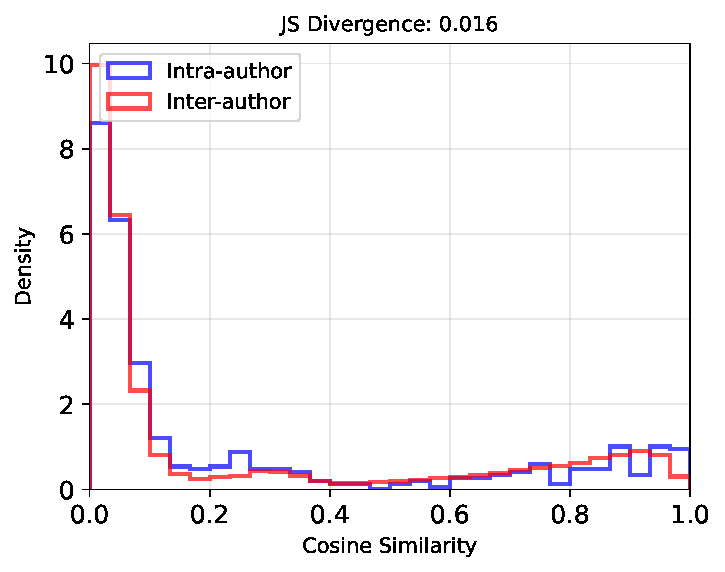}
\caption{\revision{Intra-author (blue) vs. inter-author (red) cosine similarity distributions for misclassified samples. Substantial overlap (JSD=0.016) indicates that misattributions primarily occur when different authors write stylistically similar code.}}
\Description{Overlapping density plots of intra-author and inter-author cosine similarity distributions for misclassified samples.}
\label{fig:author-overlap}
\end{figure}

\begin{itemize}
    \item \revision{\textit{Overlapping Author Styles:} Figure~\ref{fig:author-overlap} shows the distribution of intra-author and inter-author cosine similarities for misclassified samples. The two distributions exhibit substantial overlap (JSD = 0.016), indicating that misattributed code pairs have similarity scores indistinguishable from same-author pairs. For example, in the LeetCode C++ dataset, authors frequently converge on nearly identical solutions due to: (a) algorithmic constraints imposed by competitive programming problems, (b) language idioms that encourage specific patterns (e.g., STL usage), and (c) shared educational backgrounds leading to similar coding habits. Manual inspection of confused author pairs reveals cases where both authors use identical variable naming conventions (e.g., \texttt{i} and \texttt{j} for loops; \texttt{ans} for results), similar whitespace patterns, and equivalent control flow structures.}
    \item \revision{\textit{Short Code Snippets:} Figure~\ref{fig:length-brackets-performance} demonstrates a strong relationship between snippet length and attribution accuracy across models. \CodeBERT{} (Figure~\ref{fig:codebert-length}) shows accuracy increasing from 0.45 for snippets under 50 tokens to 0.64 for snippets over 1000 tokens---a 42\% improvement. \DeepSeek{} (Figure~\ref{fig:deepseek-length}) exhibits an even more pronounced pattern, with accuracy rising from 0.50 for the shortest snippets ($\leq$50 tokens) to near-perfect performance (1.0) for snippets exceeding 2000 tokens---a 100\% improvement. This trend is consistent across all models, though the magnitude varies: smaller models (\CodeBERT{}, \GraphCodeBERT{}) show gradual improvement (0.45→0.64), while larger models (\DeepSeek{}, \CodeLlama{}) exhibit steeper gains (0.50→1.0). Short snippets lack sufficient stylometric signals—they often contain only basic syntactic structures such as variable declarations, simple conditionals, or single-line operations that are nearly identical across authors regardless of individual style. For example, a 30-token snippet might consist solely of \texttt{int result = 0; for(int i=0; i$<$n; i++) result += arr[i];}---a pattern ubiquitous across authors with minimal stylistic variation. As snippet length increases, more distinguishing features emerge: function decomposition strategies, error handling patterns, commenting habits, naming convention consistency, and algorithmic approach diversity. (38\% of misattributions)}
\end{itemize}

\revision{\subsection{Why do \LLM{}s outperform \PbNN{}?}}
\revision{Language models significantly outperform \PbNN{} in code authorship attribution due to their pre-trained embeddings that capture rich semantic and syntactic code patterns learned from massive code corpora. Unlike \PbNN{}, which learns embeddings from scratch during training on limited labeled data, \LLMsS{} leverage transfer learning from large-scale pre-training, enabling them to extract more discriminative authorship features.}

\revision{To empirically validate this advantage, we analyzed the quality of learned representations by measuring the separability between intra-author (code samples from the same author) and inter-author (code samples from different authors) similarity distributions (the methodology is discussed in Section~\ref{sec:rq2-methodology}). For each model, we computed cosine similarity between code embeddings and measured the \JSD{} between the intra-author and inter-author similarity distributions. Higher \JSD{} indicates better author discrimination capability, as it reflects clearer separation between the two distributions.}

\revision{As shown in Figure~\ref{fig:intra-inter-sim-dist}, \DeepSeek{} achieves a \JSD{} of 0.992 (\GithubJava{}) and 0.666 (\LeetCode{}), compared to \PbNN{}'s 0.924 and 0.181, respectively. \DeepSeek{} shows clear bimodal separation with intra-author similarities clustered near 1.0 and inter-author similarities near 0.0, while \PbNN{} exhibits substantial overlap between distributions. This superior separability directly translates to better classification performance, explaining \LLMsS{}' advantage in code authorship attribution. We also observe the same trend in other datasets: \PbNN{}'s \JSD{} is lower than that of the other models.}

\revision{\subsection{Why are smaller \LLMsS{} more robust, and why are \LLMsS{} less vulnerable to adversarial attacks?}}

\revision{\textbf{Mechanistic Analysis of Adversarial Robustness}: To understand why smaller LMs demonstrate greater robustness to adversarial attacks compared to larger models and traditional ML/DL approaches, we conducted a mechanistic analysis using \xaiMethod{}~\cite{sundararajan2017axiomatic}. Our goal was to determine whether robustness differences stem from varying attention patterns across AST token categories. We applied the following procedure to analyze each model's decision-making process:}

\begin{enumerate}
    \item  \revision{ \textit{Token-level Attribution}: For all correctly classified samples in the test set, we computed IG attribution scores for each token, quantifying its contribution to the model's prediction.}
    
    \item \revision{\textit{Category Mapping}: Each token was mapped to one of 14 AST categories based on tree-sitter parsing: Comments, Identifiers, Keywords, Operators, Punctuation, Literals, Types, Modifiers, Declarations, Statements, Expressions, Preprocessor, Patterns, and Special tokens (Table~\ref{table:ast-categories}).}
    
    \item \revision{\textit{Category-level Aggregation}: We calculated the mean attribution score for each category across all samples, providing insight into which syntactic and semantic features each model prioritizes for authorship attribution.}
    
    \item \revision{\textit{Transformation Impact Analysis}: We cross-referenced these attribution patterns with our adversarial transformation taxonomy (Table~\ref{table:trans-ast-category}), which documents which AST categories each transformation modifies.}
\end{enumerate}

\revision{This analysis enables us to test the hypothesis that models with \textit{distributed attention} across many categories are more robust to targeted perturbations than models with \textit{concentrated attention} on few categories.}

\begin{figure}[tb]
\centering
\begin{minipage}{0.48\textwidth}
    \centering
    \includegraphics[width=\textwidth]{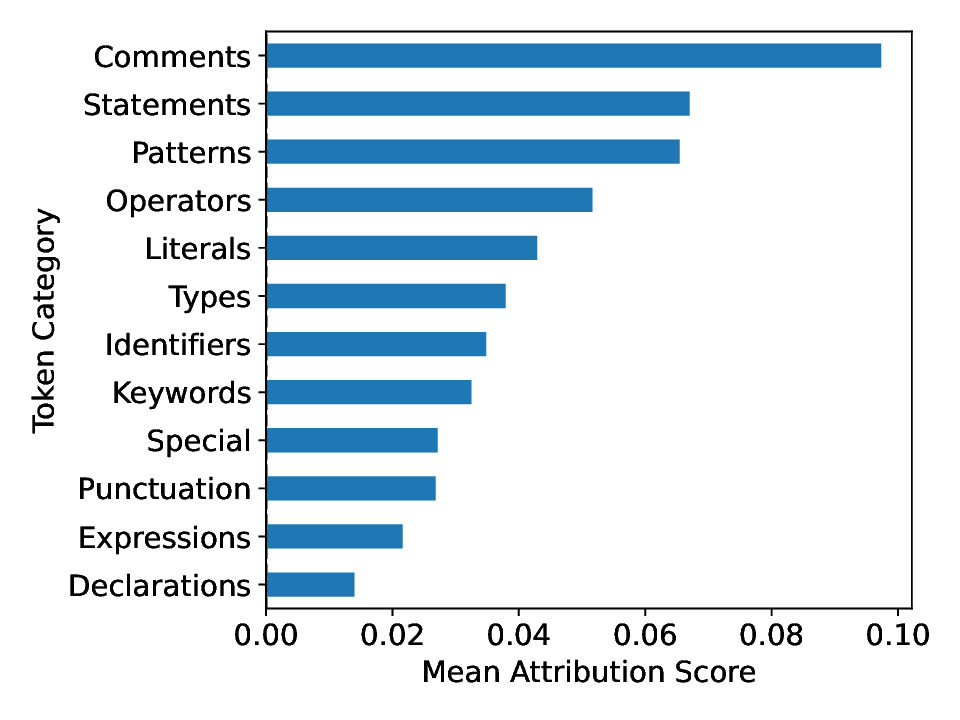}
    \caption*{(a) \CodeBERT{}}
\end{minipage}
\hfill
\begin{minipage}{0.48\textwidth}
    \centering
    \includegraphics[width=\textwidth]{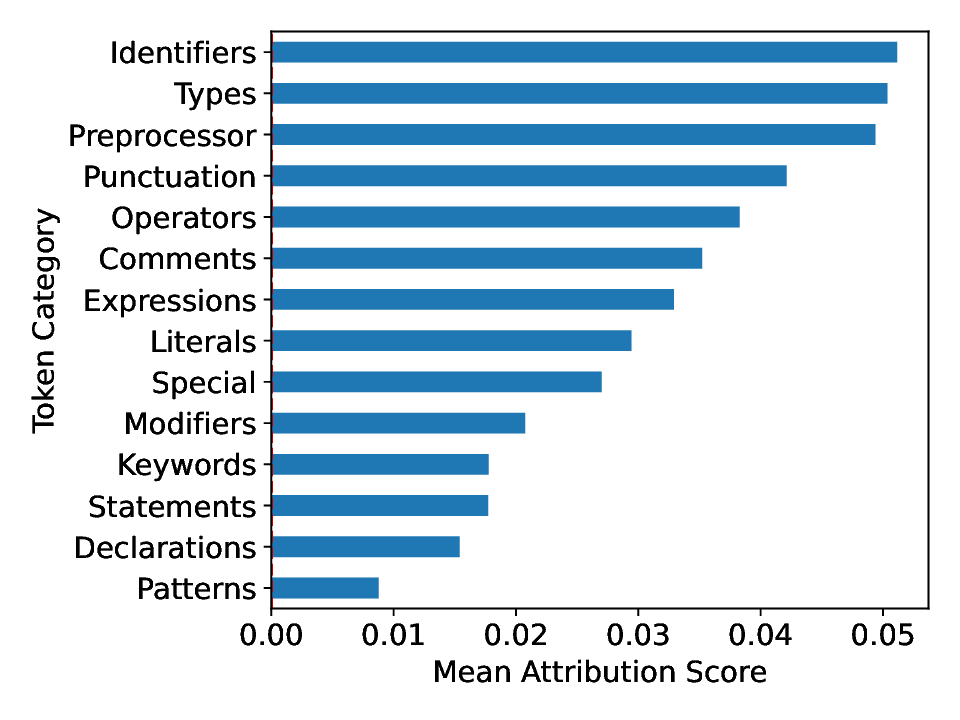}
    \caption*{(b) \ContraC{}}
\end{minipage}

\begin{minipage}{0.48\textwidth}
    \centering
    \includegraphics[width=\textwidth]{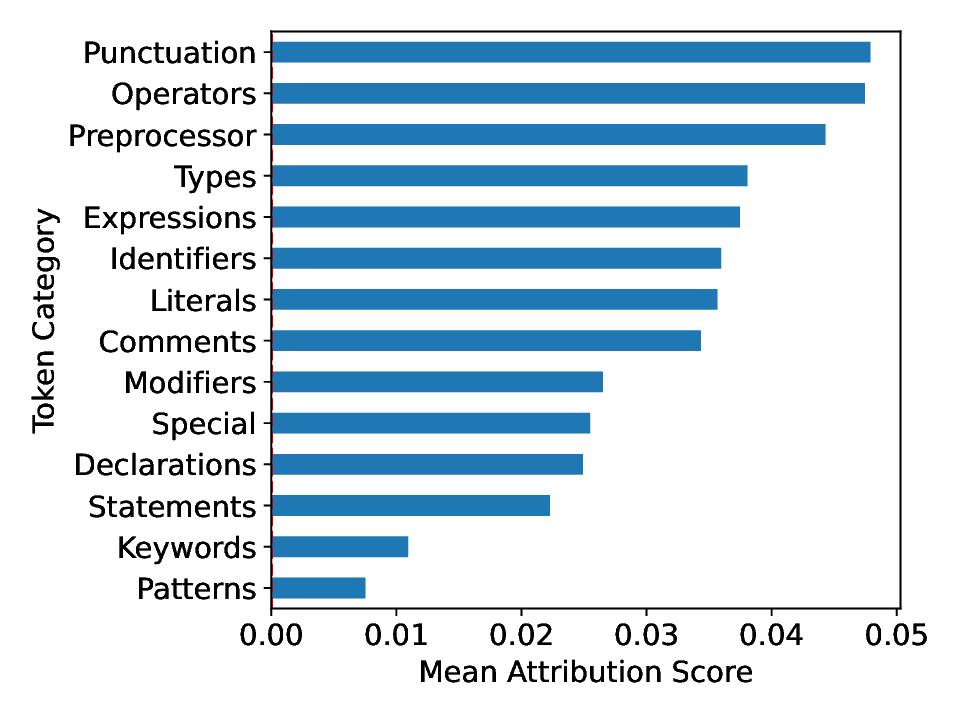}
    \caption*{(c) \ContraG{}}
\end{minipage}
\hfill
\begin{minipage}{0.48\textwidth}
    \centering
    \includegraphics[width=\textwidth]{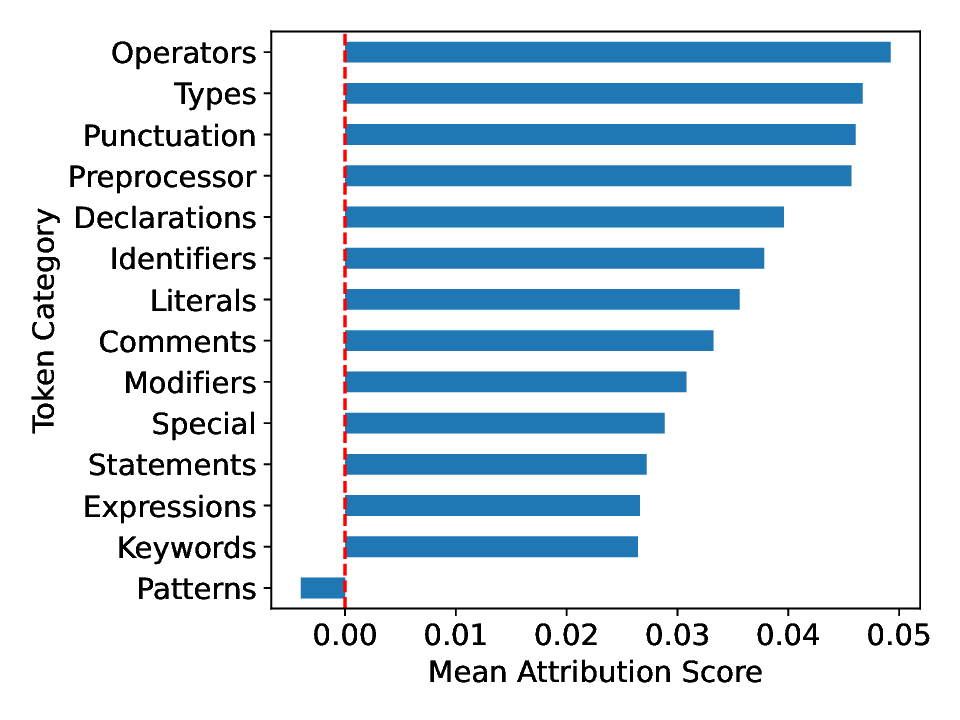}
    \caption*{(d) \GraphCodeBERT{}}
\end{minipage}
\caption{Mean attribution score of \CodeBERT{}, \ContraC{}, \ContraG{}, and \GraphCodeBERT{}}
\Description{Four bar charts showing mean attribution scores across AST categories for CodeBERT, ContraBERT\_C, ContraBERT\_G, and GraphCodeBERT, illustrating their distributed attention patterns.}
\label{fig:small-mean-attribution}
\end{figure}

\begin{figure}[tb]
\centering
\begin{minipage}{0.48\textwidth}
    \centering
    \includegraphics[width=\textwidth]{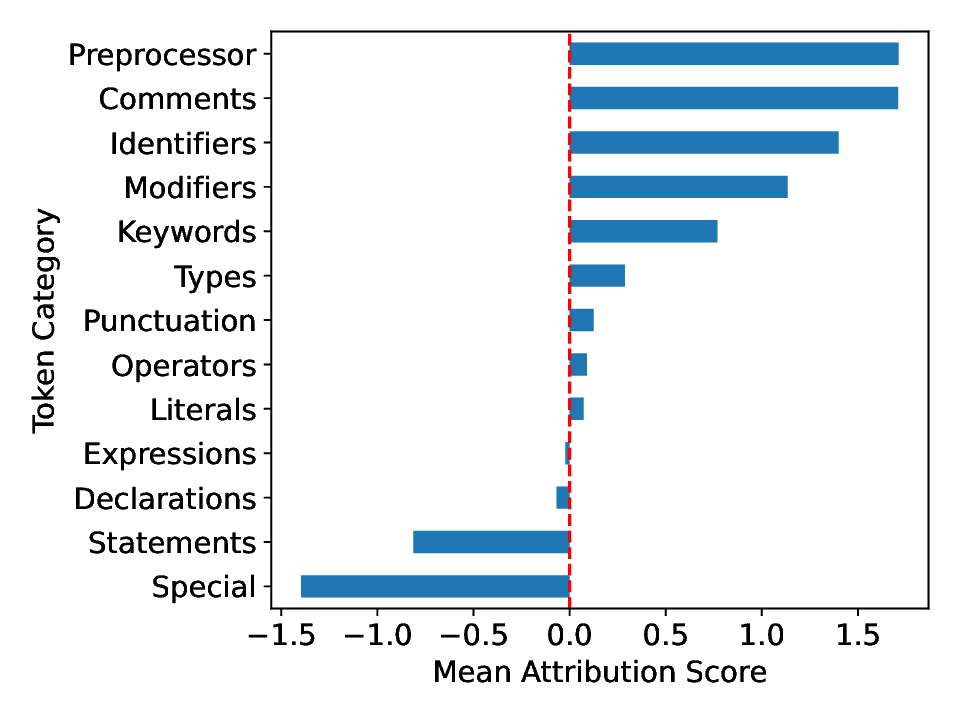}
    \caption*{(a) \PbNN{}}
\end{minipage}
\hfill
\begin{minipage}{0.48\textwidth}
    \centering
    \includegraphics[width=\textwidth]{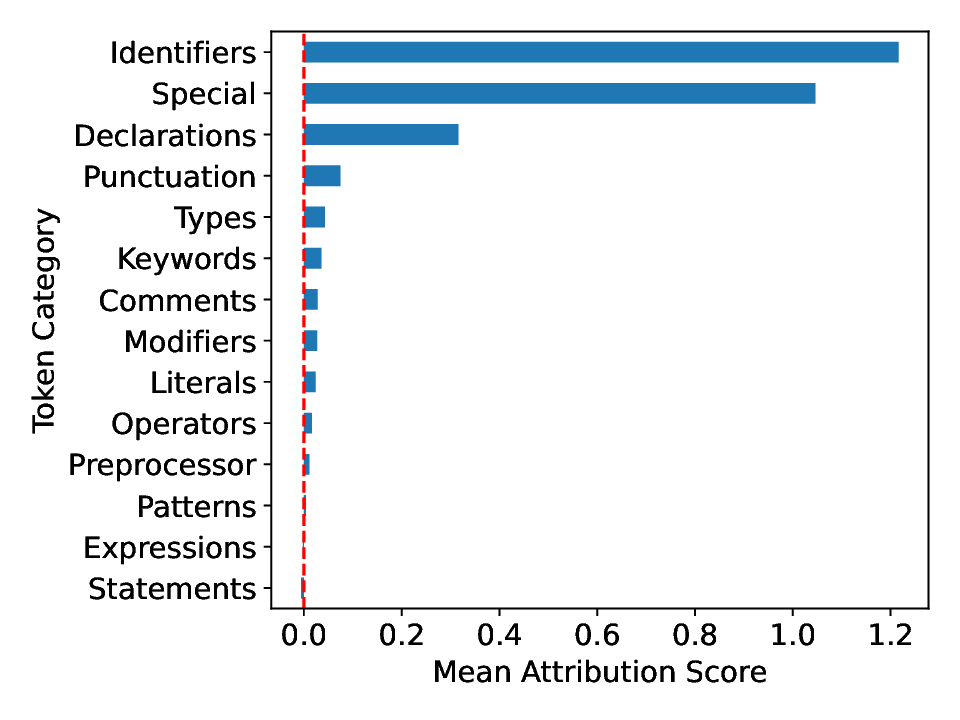}
    \caption*{(b) \UnixCoder{}}
\end{minipage}

\begin{minipage}{0.48\textwidth}
    \centering
    \includegraphics[width=\textwidth]{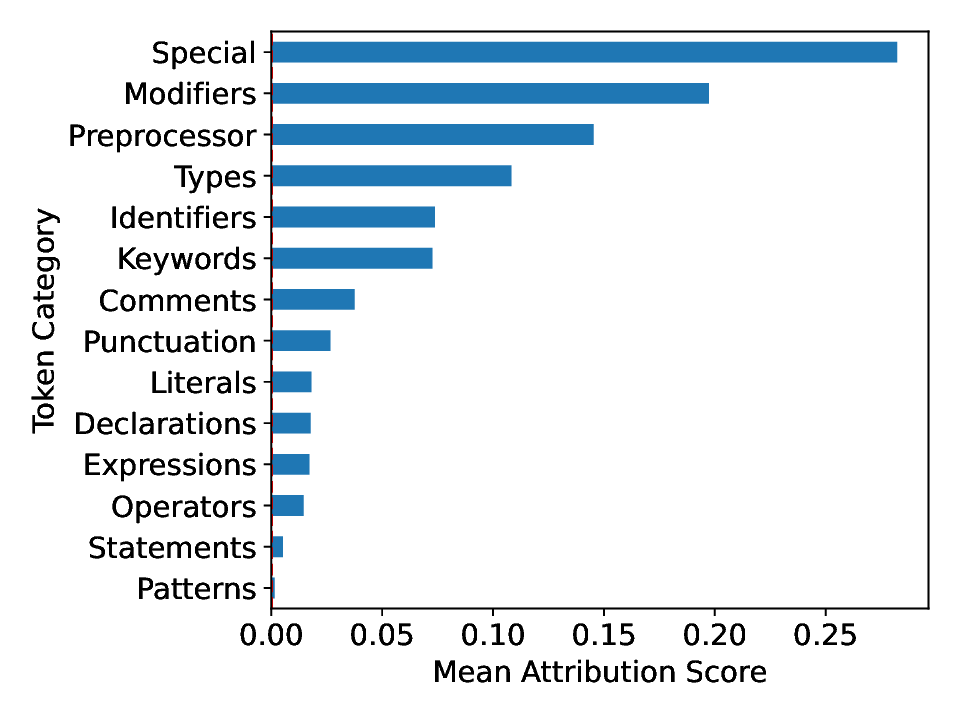}
    \caption*{(c) \DeepSeek{}}
\end{minipage}
\hfill
\begin{minipage}{0.48\textwidth}
    \centering
    \includegraphics[width=\textwidth]{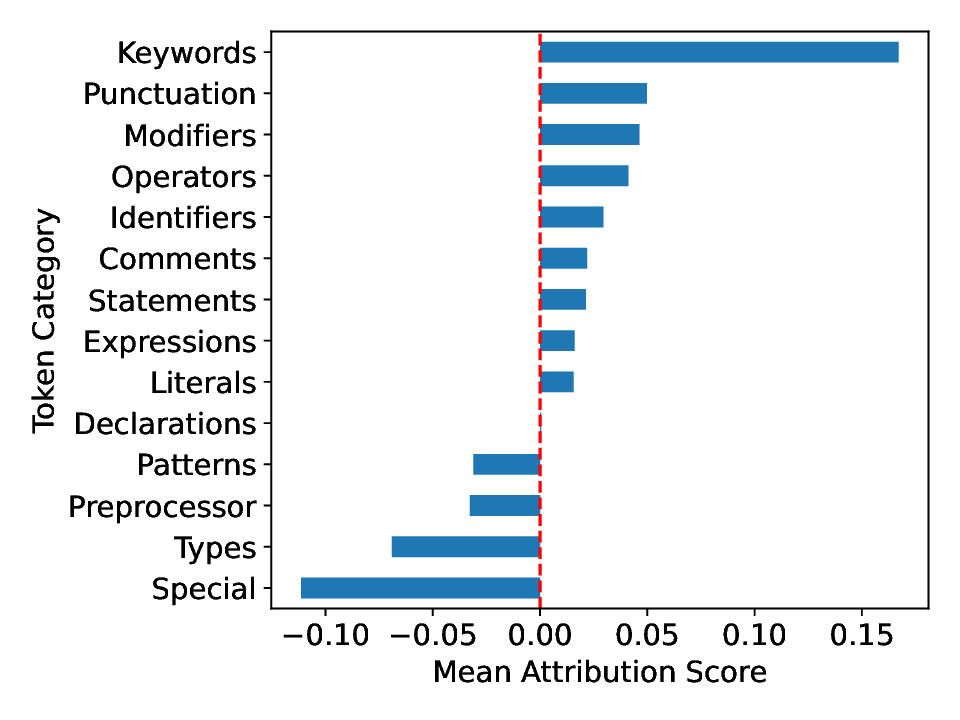}
    \caption*{(d) \CodeLlama{}}
\end{minipage}

\caption{Mean attribution score of \PbNN{}, \UnixCoder{}, \DeepSeek{} and  \CodeLlama{}}
\Description{Four bar charts showing mean attribution scores across AST categories for PbNN, UniXcoder, DeepSeek-Coder, and Code Llama, illustrating their concentrated attention patterns.}
\label{fig:large-mean-attribution}
\end{figure}

\revision{Figures~\ref{fig:small-mean-attribution} and~\ref{fig:large-mean-attribution} present the mean attribution scores across AST categories for eight models. The results reveal a clear dichotomy in attention patterns that directly explains robustness differences.}

\revision{\textbf{Smaller Models: Distributed Attention Strategy.} \CodeBERT{}, \GraphCodeBERT{}, \ContraC{}, and \ContraG{} (Figure~\ref{fig:small-mean-attribution}a-d) exhibit relatively uniform attribution distributions across most AST categories. For instance, \CodeBERT{} assigns substantial importance to Comments, Statements, Patterns, Operators, and Literals, with no single category dominating. Similarly, \GraphCodeBERT{} distributes attention across Operators, Types, Punctuation, and Preprocessor. This distributed attention pattern means these models consider \textit{holistic code structure} rather than relying on specific stylometric markers.}

\revision{\textbf{Larger Models and ML/DL: Concentrated Attention Strategy.} In contrast, larger models (\CodeLlama{}, \UnixCoder{}, \DeepSeek{}) and the traditional DL baseline (\PbNN{}) show highly concentrated attention on specific categories (Figure~\ref{fig:large-mean-attribution}a-d). \CodeLlama{} focuses predominantly on Keywords, with minimal attention to most other categories. \UnixCoder{} strongly prioritizes Identifiers and Special tokens, while largely ignoring syntactic structures. \DeepSeek{} concentrates on Special tokens and Modifiers. \PbNN{} exhibits extreme concentration on Preprocessor and Comments, with near-zero attribution to structural elements.}

\revision{\textbf{Connection to Adversarial Robustness.}
This dichotomy directly explains the observed robustness differences. As shown in Table~\ref{table:trans-ast-category}, individual adversarial transformations modify only a \textit{subset} of the 14 AST categories, and every transformation leaves at least one category untouched. The size of this footprint varies widely, from a single category to 13 (median: 11). For example:}
\begin{itemize}
    \item \revision{\textit{Remove comments} affects only 3 categories: Comments, Special, and Preprocessor}
    \item \revision{\textit{Change variable naming style} affects 6 categories: Expressions, Keywords, Punctuation, Identifiers, Operators, and Statements}
    \item \revision{\textit{Convert literals to expressions} affects 9 categories but leaves Comments, Keywords, and Modifiers unchanged}
\end{itemize}

\revision{Since our attack protocol applies only \textit{one transformation at a time}, adversarial samples preserve many original AST categories unchanged. Models with distributed attention can still leverage untouched categories for attribution, maintaining prediction accuracy. For instance, when comments are removed, \CodeBERT{} can still rely on its substantial attention to Statements, Patterns, and Operators. Conversely, models with concentrated attention are vulnerable when adversarial transformations target their focal categories. When \PbNN{} encounters code with comments removed—one of its two dominant features—it loses critical discriminative information, causing misattribution. Similarly, if identifier naming conventions are altered (affecting \UnixCoder{}'s primary focus), the model's performance degrades significantly.}

\revision{These findings reveal that adversarial robustness in CAA is not solely a function of model size or architectural sophistication, but rather depends on \textit{feature utilization strategy}. Smaller models, with limited capacity, are forced to extract signal from diverse code features, inadvertently learning more generalizable and robust representations. Larger models, with greater capacity, can achieve higher clean accuracy by specializing on a few highly discriminative features, but this specialization creates vulnerabilities.}

\revision{\subsection{Why does contrastive pre-training improve robustness?}}
\revision{
ContraBERT enhances existing pre-trained models through contrastive learning with semantic-preserving augmentations. Given original code-comment pairs (C, W), nine augmentation operators generate semantic-preserving variants (C', W') through transformations like Rename Variable, Rename Function, Reorder Statements, Insert Dead Code, and Back Translation. The model is then trained with two objectives: (1) Masked Language Modeling (MLM) to predict masked tokens, and (2) Contrastive Learning using InfoNCE loss, which maximizes similarity between original samples and their augmented variants (positive pairs) while minimizing similarity to other samples in a dynamic queue (negative pairs). This contrastive objective explicitly trains the model to treat semantically equivalent code variants as similar, forcing it to learn deep structural and syntactic features rather than surface-level patterns like variable names or statement ordering. 
}

\begin{figure*}
    \centering
    \includegraphics[width=0.7\textwidth]{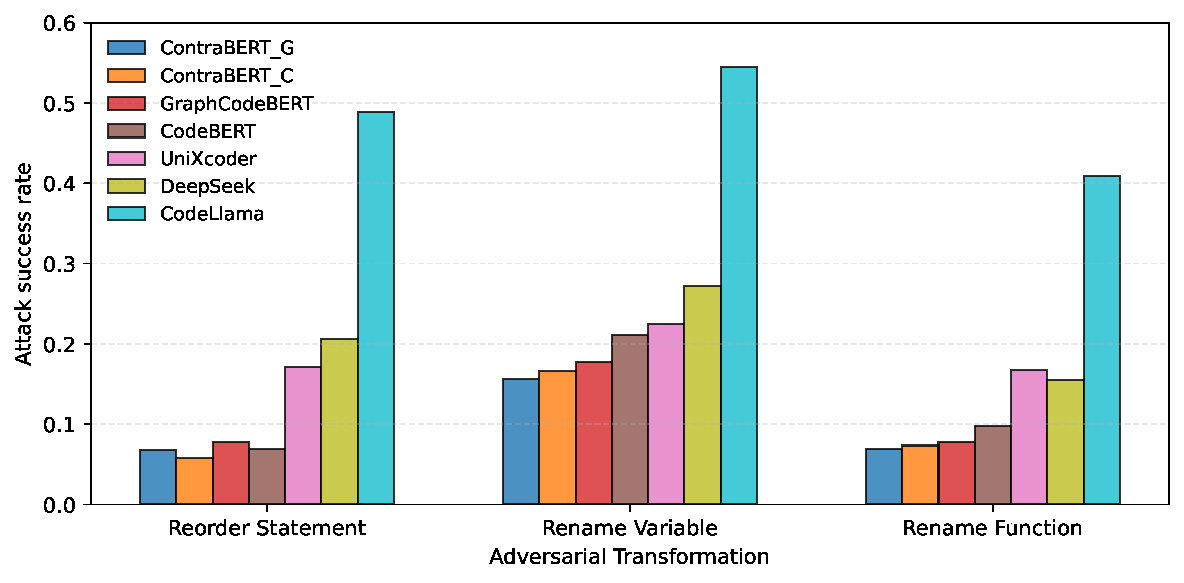}
    \caption{Model Robustness to Variable Renaming, Function Renaming, and Statement Reordering Attacks}
    \Description{Grouped bar chart comparing attack success rates of variable renaming, function renaming, and statement reordering across seven models.}
    \label{fig:contra-transformations}
\end{figure*}

\revision{To validate how ContraBERT's contrastive learning enhances stylometric feature learning for code authorship attribution, we compared attack success rates across three common code transformations that mirror ContraBERT's training augmentations: Rename Variable, Rename Function, and Reorder Statement. These semantic-preserving transformations directly test whether the model has learned to rely on stable stylometric features (syntax and structure) rather than surface-level patterns (identifier names and statement order). We evaluated seven models: \CodeLlama{}, \DeepSeek{}, \UnixCoder{}, \CodeBERT{}, \GraphCodeBERT{}, \ContraC{} (based on \CodeBERT{}), and \ContraG{} (based on \GraphCodeBERT{}).} \revision{Figure~\ref{fig:contra-transformations} presents the attack success rates across all models and transformations. ContraBERT models demonstrate substantially lower vulnerability to identifier renaming and statement reorder attacks, confirming that the contrastive objective enhances stylometric robustness.}

\section{Related Work}

Abuhamad~\etal{}~\cite{abuhamad2018large,abuhamad2021large} utilized both ML and DL to address \CAAS{}. They employed Recurrent Neural Networks (RNNs) to extract TF-IDF-based stylometric features and used a Random Forest Classifier (RFC) for authorship attribution. In a separate study, they applied Convolutional Neural Networks for the same purpose~\cite{abuhamad2019code}. Alsulami~\etal{}~\cite{alsulami2017source} proposed using an LSTM~\cite{sherstinsky2020fundamentals} network to extract relevant features from the AST representation of programmers' source code, followed by a SoftMax Layer for authorship attribution. Although their study reported high accuracy, it did not assess the model's resilience against adversarial attacks. Bogomolov~\etal{}~\cite{bogomolov2021authorship} introduced a Path-based Neural Network (\PbNN{}) for authorship attribution by modifying the \textit{code2vec} neural network architecture, focusing on the relative term frequencies of tokens and paths in the abstract syntax tree (AST). Burrows~\etal{}~\cite{burrows2014comparing,burrows2007source} used n-gram features extracted from source code to train a Support Vector Machine (SVM) for the same purpose. These studies also incorporated various manual feature engineering techniques to extract stylometric features such as byte-level information and program dependence graphs from source code~\cite{frantzeskou2006effective,caliskan2015anonymizing,ullah2019source}. Despite reporting high accuracy, these studies did not examine their models' performance under adversarial attack.

Addressing the gap in evaluating resistance to adversarial attacks, we found several studies that did assess this aspect. Quiring~\etal{}~\cite{quiring2019misleading} designed an adversarial attack strategy for \CAAS{} using Monte-Carlo tree search, which dramatically reduced the model's accuracy from 88\% to 1\%. Other studies have also shown that automated \CAAS{} models are susceptible to adversarial attacks~\cite{mcknight2018style,simko2018recognizing,liu2021practical,matyukhina2019adversarial,Ropgen}. In this study, we follow a similar methodology to perform adversarial attacks on \LLMsS{}.

Although \LLMsS{} have demonstrated promising results in various SE tasks such as code completion~\cite{raychev2014code}, code summarization~\cite{ahmed2024automatic}, code translation~\cite{pan2023understanding}, and bug triaging~\cite{dipongkor2023comparative}, their application in \CAAS{} has been relatively unexplored.  
Yang~\etal{}~\cite{yang2022natural} introduced \textit{ALERT}, a method that generates adversarial samples by renaming identifiers. They then fine-tuned and targeted \CodeBERT{} and \GraphCodeBERT{} in their attacks, affecting Vulnerability Prediction, Clone Detection, and Authorship Attribution. However, their approach to perturbing code, limited to renaming identifiers, is considerably less extensive than our 28 transformation rules. Additionally, while \textit{ALERT} tested \LLMsS{} performance using monolingual datasets, our evaluations utilize multilingual datasets, offering a more comprehensive analysis of \LLMsS{} under varied conditions.

A small portion of this work was previously published as an extended abstract for an ACM SRC competition~\cite{atish:src}. This presented preliminary results, focusing on encoder-only models with parameter sizes in the hundreds of millions. In the SRC paper, no hyper-parameter fine-tuning was performed, the experiments reported accuracy and adversarial success rates on a more limited version of our \LeetCode{} dataset for only a single model (\UnixCoder{}), there was no comparison to state-of-the-art (SOTA) baselines, and finally, interpretability techniques were not applied.

\section{Threats to Validity}

It is known that language models like \GPT{} provide non-deterministic outputs and may produce slightly different adversarial samples upon reproduction. To mitigate this threat, we use a low temperature (0), as suggested by~\cite{huang2024can}.

Another potential threat to reproducibility arises from our reliance on \GPT{}—a proprietary tool—for adversarial sample generation, rather than open-source alternatives like Llama or DeepSeek. While we attempted to use Llama, most of its adversarial samples were rejected by \LeetCode{}'s submission system, which is crucial for verifying the semantic correctness of adversarial transformations. DeepSeek provides free access through its web interface, but its API is paid, making it impractical for generating adversarial samples for 114,016 code snippets. We instead use GPT-4o's API, which is free through the GitHub Copilot plugin. Thus, researchers can reproduce our results using GitHub Copilot's plugin.

A further validity threat is that adversarial samples may reflect \GPT{}'s stylistic tendencies rather than those of a real human attacker. To mitigate this, we conduct a human evaluation of adversarial samples that significantly impact model performance. These samples, verified by our parser and \LeetCode{}'s submission system, cover all transformation rules. For this human evaluation, we followed a setup similar to that of Yang~\etal{}~\cite{yang2022natural} for evaluating the naturalness of adversarial samples.

\revision{We employ a two-step evaluation process to assess adversarial sample quality. First, one of the authors manually verifies a statistically representative sample (95\% confidence level, n=370), assessing whether: (i) \GPT{} followed the supplied transformation prompt, (ii) the adversarial code contains extraneous changes, and (iii) the transformed code remains plausible, readable, and free from obfuscation or junk code. Then, four PhD students in software engineering independently evaluate disjoint subsets of adversarial samples using the same criteria.}

\revision{\textbf{Evaluation Protocol.} Each evaluator assessed samples on three dimensions:}
\revision{\begin{itemize}
    \item \textit{Transformation adherence}: Whether \GPT{} correctly applied the specified transformation
    \item \textit{Code cleanliness}: Absence of extraneous modifications beyond the intended transformation
    \item \textit{Plausibility}: Whether the code appears human-written, rated on a 3-point scale:
    \begin{itemize}
        \item \textbf{Plausible}: Indistinguishable from typical human-written code
        \item \textbf{Moderately Plausible}: Somewhat realistic but contains minor artifacts
        \item \textbf{Not Plausible}: Contains obvious synthetic or unnatural patterns
    \end{itemize}
\end{itemize}}

\revision{
After independent reviews, evaluators discussed disagreements to reach consensus through negotiated agreement. Inter-rater agreement between the first author and the four PhD students was high, with agreement rates of 90\%, 97\%, 92\%, and 96\% respectively (mean: 93.75\%), indicating reliable evaluation.}

\revision{\textbf{Findings.}} \revision{Through this process, all evaluators agreed that \GPT{} follows the supplied prompts and that our parser effectively verifies transformations. Additionally, evaluators noted that \GPT{} tends to remove comments while applying transformations. To prevent unintended bias, we exclude such cases from the adversarial attack experiments.}

\revision{Regarding plausibility, the evaluators confirmed that the generated adversarial samples are generally realistic. However, samples involving \textit{add print/log statements} transformations received lower plausibility ratings: while 96.57\% were rated as ``Plausible,'' 3.43\% were classified as ``Moderately Plausible'' after consensus review. Manual inspection revealed these cases involved excessive logging statements (e.g., more than five print statements in the snippets), which appeared unnatural. Importantly, even these ``Moderately Plausible'' samples remained syntactically correct and executable---they simply exhibited logging density atypical of production code.}

\revision{All other transformation types (variable renaming, code reordering, comment modification, etc.) achieved 100\% ``Plausible'' ratings, indicating high realism across diverse adversarial perturbations. The 3.43\% of moderately plausible samples represent a minor limitation but do not significantly affect our conclusions for the following two reasons.}

\revision{
\begin{enumerate}
    \item These cases constitute a small fraction of total adversarial samples.
    \item They remain syntactically valid and pass LeetCode automated testing, demonstrating functional correctness.
\end{enumerate}
}

Our adversarial robustness analysis shows that larger models (e.g., \CodeLlama{}, \DeepSeek{}) exhibit higher vulnerability to adversarial attacks. This pattern may indeed reflect a tendency of larger models to overfit to specific patterns during fine-tuning, making them less robust to perturbations. However, our results also suggest that robustness is influenced not only by model size but also by training schemes and architectures (e.g., contrastive learning in ContraBERT models). In addition, we did take general measures to combat overfitting, including K-Fold cross-validation, early stopping, and dropout regularization during fine-tuning to reduce neuron co-adaptation. These combined strategies should reduce, though not entirely eliminate, the risk of overfitting.

\minorrevision{We also acknowledge that rule-based transformation tools offer higher efficiency and full determinism for syntactic 
transformations compared to our \GPT{}-based approach. However, rule-based tools are typically language-specific and 
support a narrower range of transformations. Our \GPT{}-based pipeline trades efficiency for breadth and 
language-agnosticism, enabling 28 transformation rules across multiple programming languages in a unified framework. 
We consider the efficiency gap a limitation, and future work could explore hybrid approaches that combine rule-based
tools for common syntactic transformations with \LLM{}-based generation for higher-level structural changes.}

\minorrevision{Finally, we acknowledge that our baseline comparisons rely primarily on \PbNN{}~\cite{bogomolov2021authorship}, a 2021 model. 
While one more recent CAA baseline exists~\cite{Ropgen}, replicating it within our experimental setup proved challenging due to 
language specificity and architectural constraints. We consider this a limitation of our study. Future work should 
incorporate more recent baselines as they become available with reproducible implementations, to more precisely 
contextualize the advantage of LMs for CAA.}

\section{Lessons Learned and Future Directions}

While our study primarily focuses on empirically reassessing the effectiveness, stylometric understanding, and robustness of LMs for Code Authorship Attribution (CAA), our findings suggest several concrete methods for improving future CAA systems:

\begin{enumerate}[label=\textbf{(\arabic*)}]

   \item \revision{\textbf{Ensemble Modeling.}} \revision{
   Our analysis reveals that different LMs capture orthogonal stylistic features. This suggests that an ensemble approach can enhance overall attribution accuracy and coverage. Recent work on bug triaging has demonstrated that majority voting among multiple LLMs significantly improves classification performance~\cite{kumar2024ensemble}. Building on this success, we propose extending beyond simple majority voting to weighted aggregation strategies where each model's contribution is proportional to its performance.
   }

    \item \revision{\textbf{Contrastive Pre-training.}} \revision{We observe that contrastive learning substantially improves robustness against adversarial attacks. This finding motivates the design of advanced contrastive pre-training objectives tailored for CAA. Specifically, we propose generating positive pairs from style-preserving paraphrases of the same author and hard negatives from adversarially perturbed texts or different authors with similar topics. The objective would minimize $\mathcal{L} = -\log\frac{\exp(\text{sim}(z_i, z_i^+)/\tau)}{\sum_{j} \exp(\text{sim}(z_i, z_j)/\tau)}$ where $z$ represents contextualized embeddings, enabling models to focus on robust stylistic features invariant to adversarial perturbations.}

    \item \textbf{Handling Short and Imbalanced Samples.} Our results on the LeetCode dataset highlight challenges posed by short, imbalanced code snippets with low stylometric variation. Future models could mitigate this issue using data augmentation—such as retrieving contextually similar code from large corpora—or by leveraging multi-view learning that integrates structural, lexical, and semantic representations of code.

    \item \textbf{Adversarial Robustness.} Given the susceptibility of large LMs to transformations that perturb a significant portion of the code, we advocate for training with adversarial data augmentation pipelines and syntactic regularization techniques. These strategies can improve model generalization across both benign and adversarially modified inputs.
\end{enumerate}

These insights provide a foundation for developing more accurate, generalizable, and resilient CAA systems beyond the scope of our current evaluation.

\revision{\section{Ethical Considerations}}
\revision{\subsection{Privacy and Consent}}

\revision{Code authorship attribution, while valuable for software engineering tasks, raises significant privacy concerns when applied to identify individuals. We acknowledge the following ethical considerations:}

\revision{\textbf{Privacy Risks.} CAA techniques can attribute code to individuals without their explicit consent, potentially exposing developers to unwanted identification or surveillance. In forensic and cybersecurity contexts, this capability could be misused for:
\begin{itemize}
    \item Unauthorized profiling of developers based on their coding patterns
    \item Identifying whistleblowers or anonymous contributors to sensitive projects
    \item Tracking developers across projects without consent
    \item De-anonymizing open-source contributors who prefer pseudonymity
\end{itemize}}

\revision{\textbf{Responsible Data Usage.} Our study exclusively uses publicly available datasets where code was shared under open-source licenses. However, we emphasize that deployment of CAA systems must adhere to:
\begin{itemize}
    \item Informed consent from developers whose code is analyzed
    \item Compliance with data protection regulations (e.g., GDPR, CCPA)
    \item Clear disclosure of CAA usage in terms of service agreements
    \item Respect for developers' right to pseudonymity in appropriate contexts
\end{itemize}}

\revision{\subsection{Potential Misuse and Mitigation}}

\revision{While CAA has legitimate applications in bug triaging, plagiarism detection, and copyright disputes, the technology could be misused:
\textbf{Surveillance and Discrimination.} CAA could enable:
\begin{itemize}
    \item Employer surveillance of developers' external contributions
    \item Discriminatory practices based on inferred developer characteristics
    \item Unauthorized attribution in competitive or adversarial contexts
\end{itemize}}

\revision{\textbf{Mitigation Strategies.} We recommend:
\begin{itemize}
    \item \textit{Purpose Limitation:} CAA systems should be deployed only for stated, legitimate purposes with appropriate oversight
    \item \textit{Regulatory Frameworks:} Organizations deploying CAA should establish clear policies on acceptable use, similar to biometric identification systems
    \item \textit{Transparency:} When CAA is used in production systems (e.g., plagiarism detection), users should be informed of its presence and capabilities
\end{itemize}}

\revision{\subsection{Dual-Use Nature}}

\revision{This research has a dual-use character: while it advances beneficial SE automation and forensic capabilities, it also provides techniques that could infringe on privacy. We position this work as advancing scientific understanding of code stylometry, with the expectation that practitioners will deploy these methods responsibly and within appropriate legal and ethical frameworks.}

\section{Conclusion}

In this paper, we performed the first large-scale, systematic study of applying transformer-based \LLMsS{} to the task of \textit{Code Authorship Attribution}. Our investigation included seven different \LLMsS{} across multiple datasets covering diverse languages and coding styles. We also conducted an in-depth analysis of these models' robustness to adversarial attacks. Our results show that \LLMsS{} \textit{generally excel at CAA} when properly fine-tuned, with \textit{larger models} like \CodeLlama{} and \DeepSeek{} providing superior performance on multilingual datasets that exhibit significant variance. Surprisingly, even \textit{smaller models} such as \CodeBERT{} and \GraphCodeBERT{} often perform competitively, especially in structured, monolingual scenarios. We also found that \LLMsS{}—particularly those pre-trained with contrastive objectives—are substantially \textit{more robust to adversarial perturbations} compared to earlier neural techniques.

\section{Artifacts}
We make all of our code and data available in an online appendix~\cite{online-appendix,zenodo} to facilitate the replication and reproducibility of our work, and to encourage future research on adapting transformer-based \LLMsS{} to the task of code stylometry.
\begin{acks}
This work is supported in part by NSF grants CCF-2423813. Any opinions,
findings, and conclusions expressed herein are the authors' and
do not necessarily reflect those of the sponsors. We would also like to thank the anonymous TOSEM reviewers who helped to greatly improve the quality of this manuscript.
\end{acks}

\bibliographystyle{ACM-Reference-Format}
\bibliography{references}
\clearpage

\appendix
\revision{\section{Adversarial Rules as Prompts}
\label{sec-adv-prompts-rules}
\space \textbf{A. Statement Transformation}

\begin{enumerate}
    \item Given the code snippet below, find instances where single line variable declarations can be split into multiple lines, and modify the code to use the multiple line declarations. If there is no single line declaration, just return `NA'. Otherwise, return the modified code only.
    \item Given the code snippet below, find statements where variables are declared across multiple lines, and modify the code to put those declarations into a single line. If there is no such declaration, just return `NA'. Otherwise, return the modified code only.
    \item Given the code snippet below, find statements where the order of execution does not impact the program logic. Then swap the order of these statements. Finally, return the whole code with modifications. 
\end{enumerate}

\textbf{B. Name Transformation}

\begin{enumerate}
    \item Given the code snippet below, identify if there is a particular style for variable naming, such as camel case or snake case. If there is a style for the given variable names, then change all variable names to a different style. For example, changing all instances of the variable plus\_one to plusOne, representing a shift from snake case to camel case. If there is no style convention for variable names, just return `NA'.
    \item Given the code snippet below, identify if there is a particular style for function naming, such as camel case or snake case. If there is a style for the given function names, then change all function names to a different style. For example, changing all instances of the function plus\_one to plusOne, representing a shift from snake case to camel case. If there is no style convention for function names, just return `NA'.
\end{enumerate}

\textbf{C. Operator Transformation}

\begin{enumerate}
    \item Given the code snippet below, find all instances of relational operators and swap them. For example, ` a > b ' to ` b < a '. If there are no relational operators, just return `NA'. Otherwise, return the modified code only.
    \item Given the code snippet below, identify conditional statements and convert them into their converse-negative form. For example, ` a < b ' would be transformed into ` !(a > b) '. If there are no conditional statements, just return `NA'. Otherwise, return the modified code only.
    \item Given the code snippet below, identify and convert integer literals into expressions that represent the same literal value. For example, ` int b = 8 ' would be transformed into ` int b = 2 * 4 '. If there are no integer literals, just return `NA'. Otherwise, return the modified code only.
    \item For the code snippet below, identify and change the style of any increment or decrement operators. For example, ` i++ ' would change to ` i +=1' or ` i-{}- ' would change to ` i -= 1 '. If there are no increment or decrement operators, just return `NA'. Otherwise, return the modified code only. 
    \item For the code snippet below, identify and change the style of any increment or decrement operators. For example, ` i +=1' would change to ` i++ ' or ` i -= 1 ' would change to ` i-{}- ' . If there are no increment or decrement operators, just return `NA'. Otherwise, return the modified code only.
\end{enumerate}

\textbf{D. Data Transformation}

\begin{enumerate}
    \item Given the code snippet below, identify and convert all integer numbers into hexadecimal values. If there are no integer numbers, just return `NA'. Otherwise, return the modified code only.
    \item Given the code snippet below, convert all character literals into their corresponding ASCII values. If there are no character literals, just return `NA'. Otherwise, return the modified code only.
    \item Given the code snippet below, identify any string variables and modify their declaration to use a character array. If there are no string variables, just return `NA'. Otherwise, return the modified code only.
    \item Given the code snippet below, identify and convert between boolean literals and integer literals. For example, if a `true' value is used change this to a 1 and if a `false' value is used change this to a 0. If there are no boolean values used, just return `NA'. Otherwise, return the modified code only.
    
\end{enumerate}

\textbf{E. Loop Transformation}

\begin{enumerate}
    \item Given the code snippet below, identify and change any `for' statements into `while' statements. If there are no `for' statements, just return `NA'. Otherwise, return the modified code only. 
    \item Given the code snippet below, identify and change any `while' statements into `for' statements. If there are no `while' statements, just return `NA'. Otherwise, return the modified code only.
\end{enumerate}

\textbf{F. Control Flow Transformation}

\begin{enumerate}
    \item Given the code snippet below, identify and convert any eligible `if-else' statements to `switch' statements. If there are no `if-else' statements that can be converted into `switch' statements, just return `NA'. Otherwise, return the modified code only. 
    \item Given the code snippet below, identify and convert any eligible `switch' statements to `if-else' statements. If there are no `switch' statements that can be converted into `if-else' statements, just return `NA'. Otherwise, return the modified code only.
    \item Given the code snippet below, identify and convert any `if-else' statements that can be rewritten using the `ternary' operator. For example, ` if (a > b)\{ max = a; \} else \{max = b;\} ' would be converted into ` max = a > b ? a : b; '. If there is no such statement, just return `NA'. Otherwise, return the modified code only.
    \item Given the code snippet below, identify and convert any `if-else' statements written using the `ternary' operator to `if-else' statements that do not use the `ternary' operator. For example, ` max = a > b ? a : b; ' would be converted into ` if (a > b)\{ max = a; \} else \{max = b;\} '. If there is no such statement, just return `NA'. Otherwise, return the modified code only. 
    \item Given the code snippet below, identify any `if-else' statements and swap the statements contained within the if conditional to the else conditional. If there are no `if-else' conditional statements, just return `NA'. Otherwise, return the modified code only.
\end{enumerate}

\textbf{G. Function Transformation}

\begin{enumerate}
    \item Given the code snippet below, identify function declarations and invocations, and swap the order of the parameters for both the declarations and invocations. If there are no function declarations, just return `NA'. Otherwise, return the modified code only.
    \item Given the code snippet below, identify function declarations and invocations, and add one extra integer parameter with a default value of zero. If there are no function declarations, just return `NA'. Otherwise, return the modified code only.
    \item Given the code snippet below, identify groups of statements that could be rewritten into a function, and create a function and any necessary invocations. Finally, return the modified code Only. 
    \item Given the code snippet below, identify any function declarations, and swap the order of their declarations. If there are one or fewer function declarations, do not modify the code. If there is no such declaration, just return `NA'. Otherwise, return the modified code only.
\end{enumerate}

\textbf{H. Miscellaneous}

\begin{enumerate}
    \item Given the code snippet below, please remove all comments. If there is no comment, just return `NA'. Otherwise, return the modified code only.
    \item Given the code snippet below, please remove unused code including variables, functions, libraries. Only remove code that is never used for computational logic. If there is no unused code, just return `NA'. Otherwise, return the modified code only.
    \item Given the code snippet below, please add print/log statements at every point where a variable is initialized or its value modified. Finally, return the modified code only.
\end{enumerate}}

\begin{table}[tb]
\caption{\revision{Best Learning Rate (LR) and Batch Size (BS) for different models. BS denotes the \textit{effective} batch size; for \DeepSeek{} it is realized through gradient accumulation over a per-device micro-batch of 4. \CodeLlama{} is omitted because memory constraints permitted tuning only its learning rate.}}\label{table:hyperparams}
\setlength{\tabcolsep}{1pt}
    \centering
\begin{tabular}{l|cc|cc|cc|cc|cc|cc}
\hline
\multirow{2}{*}{Dataset} & \multicolumn{2}{c|}{\CodeBERT}      & \multicolumn{2}{c|}{\ContraC} & \multicolumn{2}{c|}{\ContraG} & \multicolumn{2}{c|}{\GraphCodeBERT} & \multicolumn{2}{c|}{\UnixCoder}     & \multicolumn{2}{c}{\DeepSeek}      \\ \cline{2-13} 
                         & \multicolumn{1}{c|}{LR}       & BS & \multicolumn{1}{c|}{LR}       & BS & \multicolumn{1}{c|}{LR}       & BS & \multicolumn{1}{c|}{LR}       & BS & \multicolumn{1}{c|}{LR}       & BS & \multicolumn{1}{c|}{LR}       & BS \\ \hline
\GCJCPP                  & \multicolumn{1}{c|}{3.0e-05} & 16 & \multicolumn{1}{c|}{3.0e-05} & 16 & \multicolumn{1}{c|}{5.0e-05} & 16 & \multicolumn{1}{c|}{5.0e-05} & 16 & \multicolumn{1}{c|}{5.0e-05} & 16 & \multicolumn{1}{c|}{5.0e-05} & 16 \\ \hline
\GCJJava                 & \multicolumn{1}{c|}{2.0e-05} & 32 & \multicolumn{1}{c|}{3.0e-05} & 16 & \multicolumn{1}{c|}{5.0e-05} & 32 & \multicolumn{1}{c|}{2.0e-05} & 16 & \multicolumn{1}{c|}{2.0e-05} & 32 & \multicolumn{1}{c|}{2.0e-05} & 16 \\ \hline
\GCJPython               & \multicolumn{1}{c|}{3.0e-05} & 16 & \multicolumn{1}{c|}{5.0e-05} & 32 & \multicolumn{1}{c|}{5.0e-05} & 16 & \multicolumn{1}{c|}{5.0e-05} & 16 & \multicolumn{1}{c|}{5.0e-05} & 32 & \multicolumn{1}{c|}{3.0e-05} & 16 \\ \hline
\GithubC                 & \multicolumn{1}{c|}{2.0e-05} & 16 & \multicolumn{1}{c|}{2.0e-05} & 16 & \multicolumn{1}{c|}{3.0e-05} & 32 & \multicolumn{1}{c|}{2.0e-05} & 16 & \multicolumn{1}{c|}{2.0e-05} & 16 & \multicolumn{1}{c|}{2.0e-05} & 16 \\ \hline
\GithubJava              & \multicolumn{1}{c|}{5.0e-05} & 32 & \multicolumn{1}{c|}{3.0e-05} & 32 & \multicolumn{1}{c|}{5.0e-05} & 32 & \multicolumn{1}{c|}{3.0e-05} & 16 & \multicolumn{1}{c|}{2.0e-05} & 32 & \multicolumn{1}{c|}{3.0e-05} & 16 \\ \hline
\LeetCode                 & \multicolumn{1}{c|}{5.0e-05} & 16 & \multicolumn{1}{c|}{5.0e-05} & 32 & \multicolumn{1}{c|}{3.0e-05} & 16 & \multicolumn{1}{c|}{3.0e-05} & 16 & \multicolumn{1}{c|}{5.0e-05} & 16 & \multicolumn{1}{c|}{2.0e-05} & 16 \\ \hline
\end{tabular}
\end{table}

\revision{\section{Details of LeetCode Dataset}
As shown in Table~\ref{tbl:leetcode-details-stat}, the dataset exhibits two primary characteristics:
\begin{enumerate}
    \item \textbf{Language imbalance}: C++ (62.19\%) and Java (24.83\%) dominate the dataset, while Python (6.77\%), JavaScript (3.79\%), C\# (2.02\%), and Ruby (0.40\%) are underrepresented.
    \item \textbf{Author imbalance}: Within each language, author contribution varies substantially, as evidenced by the high standard deviations of samples per author (ranging from 4.99 to 23.86) and the large gaps between minimum and maximum contributions. For instance, in C++, individual authors contributed between 1 and 150 samples (std: 21.05), while in C\#, the range is 1 to 56 samples with only 4 authors (std: 23.86).
\end{enumerate}}
\begin{table}[tb]
\centering
\caption{\revision{Details of the \LeetCode{} dataset. Author counts and samples-per-author (S/A) statistics are computed within each language; authors who contribute in multiple languages are counted once per language, so the author counts sum to more than the 198 distinct authors in the dataset.}}
\label{tbl:leetcode-details-stat}
\begin{tabular}{l|l|l|l|l|l}
\hline
Language   & \# of Sample (S) & \# of Author (A) & Min S/A & Max S/A & Std (S/A) \\ \hline
\cpp{}     & 2956             & 142              & 1       & 150     & 21.05     \\ \hline
\java{}    & 1180             & 87               & 1       & 82      & 13.26     \\ \hline
Python     & 322              & 29               & 1       & 109     & 21.19     \\ \hline
JavaScript & 180              & 20               & 1       & 38      & 10.09     \\ \hline
C\#        & 96               & 4                & 1       & 56      & 23.86     \\ \hline
Ruby       & 19               & 3                & 1       & 13      & 4.99      \\ \hline
\end{tabular}
\end{table}

\begin{figure}
    \centering
    \includegraphics[width=0.5\linewidth]{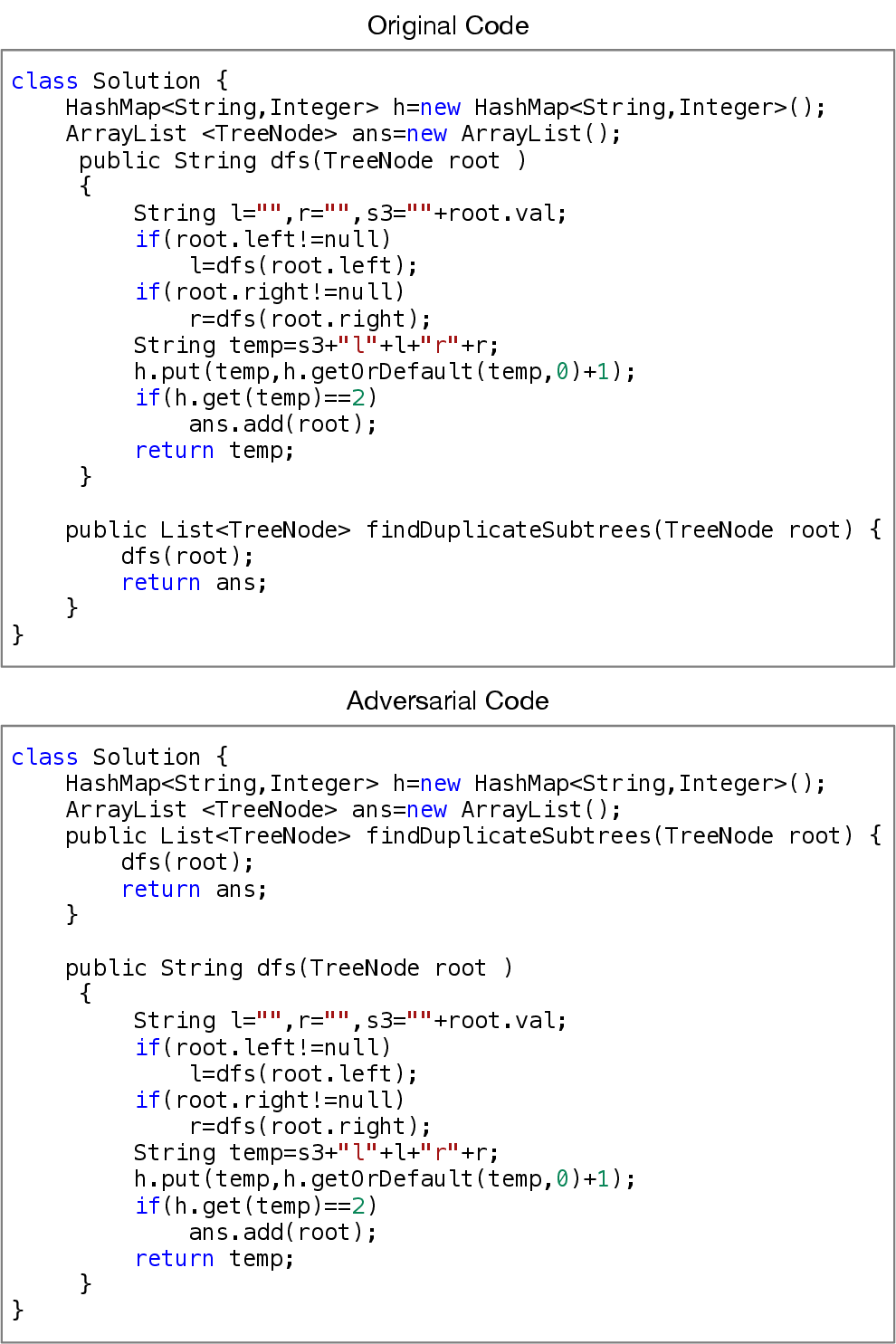}
    \caption{Adversarial Change: Swapping the order of function declarations}
    \Description{Code diff showing an adversarial transformation that swaps the order of function declarations in source code.}
    \label{fig:category-h}
\end{figure}

\begin{figure}
    \centering
    \includegraphics[width=0.8\linewidth]{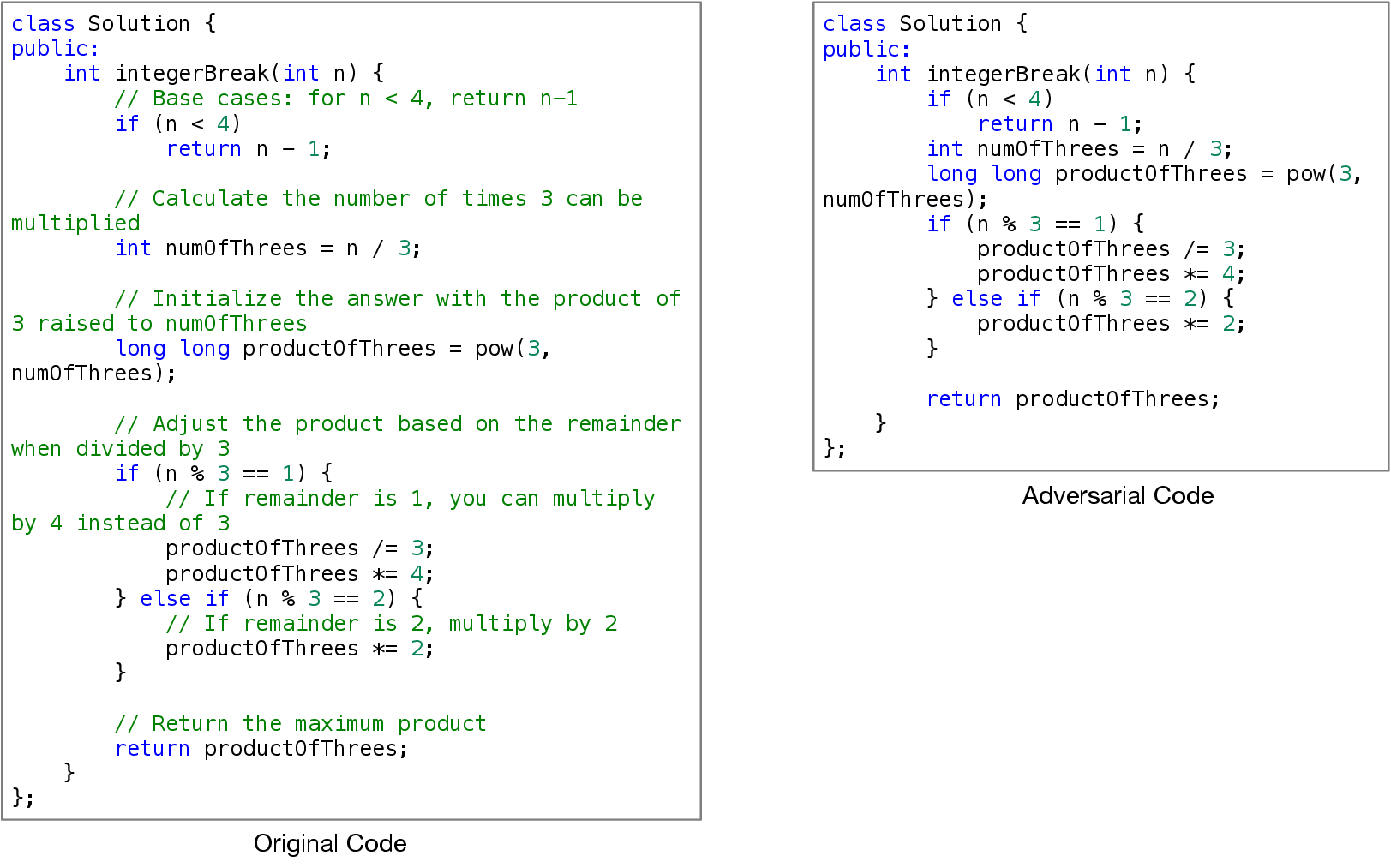}
    \caption{Adversarial Change: Removing comments}
    \Description{Code diff showing an adversarial transformation that removes all comments from source code.}
    \label{fig:category-g}
\end{figure}

\begin{table}[tb]
\centering
\caption{AST Node Categories and Representative Types}
\label{table:ast-categories}
\begin{tabular}{l|p{8cm}}
\hline
\textbf{AST Category} & \textbf{AST Node Types} \\ \hline
Comments (CMT) & comment, line\_comment, block\_comment, documentation\_comment \\ \hline
Identifiers (IDN) & identifier, field\_identifier, type\_identifier, property\_identifier, variable\_name, function\_name, method\_name \\ \hline
Keywords (KWD) & if, else, for, while, return, class, def, function, import, try, catch, switch, break, continue \\ \hline
Operators (OPR) & +, -, *, /, ==, !=, \&\&, ||, =, +=, ++, -{}-, \%, **, binary\_operator, unary\_operator \\ \hline
Literals (LIT) & integer, float, string, boolean\_literal, true, false, null, string\_literal, number\_literal \\ \hline
Types (TYP) & int, float, double, boolean, void, string, primitive\_type, array\_type, generic\_type \\ \hline
Declarations (DCL) & function\_definition, method\_declaration, class\_definition, variable\_declarator, parameter\_list \\ \hline
Statements (STM) & if\_statement, for\_statement, while\_statement, return\_statement, block, expression\_statement \\ \hline
Expressions (EXP) & binary\_expression, assignment\_expression, call, field\_expression, parenthesized\_expression \\ \hline
Modifiers (MOD) & public, private, protected, static, final, const, abstract, virtual, async \\ \hline
Punctuation (PUN) & \{, \}, (, ), [, ], ;, ., {,}, ::, and other delimiter tokens \\ \hline
Preprocessor (PRP) & preproc\_include, preproc\_def, preproc\_ifdef, preproc\_function\_def, system\_lib\_string, \#include, \#define \\ \hline
Patterns (PAT) & pattern, tuple\_pattern, list\_pattern, dictionary\_pattern, identifier\_pattern, wildcard\_pattern \\ \hline
Special (SPL) & tokenizer control tokens (e.g., [CLS], [SEP], [PAD], [MASK]) and subword-continuation markers \\ \hline
\end{tabular}
\end{table}
\begin{table}[tb]
\centering
\caption{Adversarial Transformation vs Affected AST Category}
\label{table:trans-ast-category}
\small
\setlength{\tabcolsep}{0.4pt}
\begin{tabular}{l|c|p{7cm}}
\hline
\textbf{Category}              & \textbf{Transformation}                      & \textbf{Categories Affected (CA)}                                                                                                                 \\ \hline
\multirow{3}{*}{Miscellaneous} & Remove comments                              & CMT, SPL, PRP                                                                                                                   \\ \cline{2-3} 
                               & Remove unused code                           & TYP, EXP, LIT, KWD, DCL, MOD, PAT, PUN, IDN, OPR, STM, PRP          \\ \cline{2-3} 
                               & Add print/log statements                     & TYP, EXP, LIT, KWD, CMT, DCL, MOD, PUN, IDN, OPR, STM, SPL, PRP \\ \hline
\multirow{3}{*}{Statement}     & Split variable declarations                  & TYP, EXP, LIT, KWD, DCL, MOD, PAT, PUN, IDN, OPR, STM                        \\ \cline{2-3} 
                               & Merge variable declarations                  & TYP, EXP, LIT, KWD, DCL, MOD, PAT, PUN, IDN, OPR, STM                        \\ \cline{2-3} 
                               & Reorder statements                           & TYP, EXP, LIT, KWD, CMT, DCL, MOD, PUN, IDN, OPR, STM                        \\ \hline
\multirow{2}{*}{Naming}        & Change variable naming style                 & EXP, KWD, PUN, IDN, OPR, STM                                                                            \\ \cline{2-3} 
                               & Change function naming style                 & IDN                                                                                                                       \\ \hline
\multirow{5}{*}{Operator}      & Swap relational operators                    & TYP, EXP, LIT, KWD, DCL, PUN, IDN, OPR, STM, SPL                                    \\ \cline{2-3} 
                               & Convert to converse-negative                 & EXP, LIT, KWD, PUN, IDN, OPR, STM, SPL, PRP                                           \\ \cline{2-3} 
                               & Convert literals to expressions              & TYP, EXP, LIT, DCL, PUN, IDN, OPR, STM, SPL                                              \\ \cline{2-3} 
                               & Change ++ to +=1 & EXP, LIT, KWD, PUN, IDN, OPR, STM, SPL                                                         \\ \cline{2-3} 
                               & Change +=1 to ++ & EXP, LIT, KWD, PUN, IDN, OPR, STM                                                                  \\ \hline
\multirow{4}{*}{Data}          & Convert integers to hexadecimal              & TYP, EXP, LIT, DCL, PUN, IDN, OPR, SPL                                                          \\ \cline{2-3} 
                               & Convert characters to ASCII values           & TYP, EXP, LIT, PUN, IDN, OPR, SPL                                                                        \\ \cline{2-3} 
                               & Convert strings to character arrays          & TYP, EXP, LIT, KWD, DCL, MOD, PUN, IDN, OPR, STM, SPL                         \\ \cline{2-3} 
                               & Convert booleans to integers                 & TYP, EXP, LIT, DCL, PUN, IDN, OPR                                                                   \\ \hline
\multirow{2}{*}{Loop}          & Convert for to while loops                   & TYP, EXP, LIT, KWD, DCL, MOD, PAT, PUN, IDN, OPR, STM, SPL               \\ \cline{2-3} 
                               & Convert while to for loops                   & TYP, EXP, LIT, KWD, DCL, PUN, IDN, OPR, STM                                             \\ \hline
\multirow{5}{*}{Control Flow}  & Convert if-else to switch                    & TYP, EXP, LIT, KWD, DCL, MOD, PUN, IDN, OPR, STM, SPL                         \\ \cline{2-3} 
                               & Convert switch to if-else                    & TYP, EXP, KWD, DCL, PUN, IDN, OPR, STM                                                       \\ \cline{2-3} 
                               & Convert if-else to ternary                   & TYP, EXP, LIT, KWD, DCL, MOD, PAT, PUN, IDN, OPR, STM, SPL               \\ \cline{2-3} 
                               & Convert ternary to if-else                   & TYP, EXP, LIT, KWD, DCL, PAT, PUN, IDN, OPR, STM, SPL                          \\ \cline{2-3} 
                               & Swap if-else branches                        & EXP, LIT, KWD, CMT, PUN, IDN, OPR, STM, SPL                                               \\ \hline
\multirow{4}{*}{Function}      & Swap function parameters                     & TYP, EXP, LIT, KWD, DCL, MOD, PUN, IDN, OPR, STM, SPL, PRP           \\ \cline{2-3} 
                               & Add default parameter                        & TYP, EXP, LIT, KWD, DCL, MOD, PUN, IDN, OPR, STM, PRP                    \\ \cline{2-3} 
                               & Extract statements into function             & TYP, EXP, LIT, KWD, DCL, MOD, PAT, PUN, IDN, OPR, STM, SPL               \\ \cline{2-3} 
                               & Reorder function declarations                & TYP, EXP, LIT, KWD, DCL, MOD, PUN, IDN, OPR, STM, PRP                    \\ \hline
\end{tabular}

\footnotesize
CMT=Comments, IDN=Identifiers, KWD=Keywords, OPR=Operators, LIT=Literals, TYP=Types, DCL=Declarations, STM=Statements, EXP=Expressions, MOD=Modifiers, PAT=Patterns, PUN=Punctuation, SPL=Special, PRP=Preprocessor
\end{table}

\begin{table}[tb]
\centering
\caption{LoRA configuration vs \CodeLlama{}'s performance}
\label{tbl-lora-config}
\begin{tabular}{l|l|l|l|l|l}
\hline
\textbf{rank} & \textbf{alpha} & \textbf{accuracy} & \textbf{precision} & \textbf{recall} & \textbf{f1} \\ \hline
32            & 64             & 0.68              & 0.65               & 0.68            & 0.65        \\ \hline
16            & 32             & 0.69              & 0.66               & 0.69            & 0.66        \\ \hline
8             & 16             & 0.68              & 0.65               & 0.68            & 0.65        \\ \hline
\end{tabular}
\end{table}

\end{document}